\documentclass[twocolumn]{aastex63}
\usepackage{verbatim}
\usepackage{comment}

\makeatletter
\def\@hex@@Hex#1%
 {\if a#1A\else \if b#1B\else \if c#1C\else \if d#1D\else
  \if e#1E\else \if f#1F\else #1\fi\fi\fi\fi\fi\fi \@hex@Hex}
\makeatother
\definecolor{afcolor}{HTML}{b3443c}

\revised{\today}
\shorttitle{LyC escape fraction}
\shortauthors{Ferrara et al.}
\graphicspath{{./}{figures/}}

\begin{document}



\def\be{\begin{equation}}
\def\ee{\end{equation}}
\newcommand\code[1]{\textsc{\MakeLowercase{#1}}}
\newcommand\quotesingle[1]{`{#1}'}
\newcommand\quotes[1]{``{#1}"}
\def\gsim{\lower.5ex\hbox{\gtsima}} 
\def\lsim{\lower.5ex\hbox{\ltsima}} 
\def\gtsima{$\; \buildrel > \over \sim \;$} 
\def\ltsima{$\; \buildrel < \over \sim \;$} \def\gsim{\lower.5ex\hbox{\gtsima}} 
\def\lsim{\lower.5ex\hbox{\ltsima}} 
\def\simgt{\lower.5ex\hbox{\gtsima}} 
\def\simlt{\lower.5ex\hbox{\ltsima}}

\def\msun{{\rm M}_{\odot}}
\def\lsun{{\rm L}_{\odot}}
\def\dsun{{\cal D}_{\odot}}
\def\fsun{\xi_{\odot}}
\def\zsun{{\rm Z}_{\odot}}
\def\msunyr{\msun {\rm yr}^{-1}}
\def\gdens{\msun\,{\rm kpc}^{-2}}
\def\sfrdens{\msun\,{\rm yr}^{-1}\,{\rm kpc}^{-2}}

\def\mum{\mu {\rm m}}
\newcommand{\angstrom}{\mbox{\normalfont\AA}}
\def\cc{\rm cm^{-3}}
\def\uflux{{\rm erg}\,{\rm s}^{-1} {\rm cm}^{-2} }

\def\fdust{\xi_{d}}
\def\fesc{f_{\rm esc}}
\def\td{\tau_{sd}}
\def\Sg{$\Sigma_{g}$}
\def\S*{$\Sigma_{\rm SFR}$}
\def\Ssfr{\Sigma_{\rm SFR}}
\def\Sgas{\Sigma_{\rm g}}
\def\Sstar{\Sigma_{\rm *}}
\def\Sesc{\Sigma_{\rm esc}}
\def\Srad{\Sigma_{\rm rad}}
\def\sSFR{\rm sSFR}

\def\Dsolar{${\cal D}/\dsun$}
\def\Zsolar{$Z/\zsun$}
\def\DDsolar{\left( {{\cal D}\over \dsun} \right)}
\def\ZZsolar{\left( {Z \over \zsun} \right)}
\def\kms{{\rm km\,s}^{-1}\,}
\def\skms{$\sigma_{\rm kms}\,$}

\def\Scii{$\Sigma_{\rm [CII]}$}
\def\Sciimax{$\Sigma_{\rm [CII]}^{\rm max}$}
\def\CII{\hbox{[C~$\scriptstyle\rm II $]~}}
\def\CIII{\hbox{C~$\scriptstyle\rm III $]~}}
\def\OII{\hbox{[O~$\scriptstyle\rm II $]~}}
\def\OIII{\hbox{[O~$\scriptstyle\rm III $]~}}
\def\HH{\hbox{H$_2$}~} 
\def\HI{\hbox{H~$\scriptstyle\rm I\ $}} 
\def\HII{\hbox{H~$\scriptstyle\rm II\ $}} 
\def\CIion{\hbox{C~$\scriptstyle\rm I $~}}
\def\CIIion{\hbox{C~$\scriptstyle\rm II $~}}
\def\CIIIion{\hbox{C~$\scriptstyle\rm III $~}}
\def\CIVion{\hbox{C~$\scriptstyle\rm IV $~}}
\def\nhh{n_{\rm H2}}
\def\nhi{n_{\rm HI}}
\def\nhii{n_{\rm HII}}
\def\fhh{x_{\rm H2}}
\def\fhi{x_{\rm HI}}
\def\fhii{x_{\rm HII}}
\def\fd{f^*_{\rm diss}} 
\def\ks{\kappa_{\rm s}}

\def\cyan{\color{cyan}}
\definecolor{apcolor}{HTML}{b3003b}
\definecolor{afcolor}{HTML}{800080}
\definecolor{lvcolor}{HTML}{DF7401}
\definecolor{mdcolor}{HTML}{01abdf} 
\definecolor{cbcolor}{HTML}{ff0000}
\definecolor{sccolor}{HTML}{cc5500} 
\definecolor{sgcolor}{HTML}{00cc7a}

\title{Redshift evolution of Lyman continuum escape fraction after {JWST}}

\author[0000-0002-9400-7312]{A. Ferrara}
\email{andrea.ferrara@sns.it}
\affil{Scuola Normale Superiore,  Piazza dei Cavalieri 7, 50126 Pisa, Italy}
\author[0000-0002-7831-8751]{M. Giavalisco}
\affil{University of Massachusetts Amherst, 710 North Pleasant Street, Amherst, MA 01003-9305, USA}
\author[0000-0001-8940-6768]{L. Pentericci}
\affil{INAF - OAR, via Frascati 33, 00078 Monte Porzio Catone, Roma, Italy} 
\author[0000-0001-8940-6768]{E. Vanzella}
\affil{INAF - OAS, Via Piero Gobetti, 93/3, 40129 Bologna, Italy} 
\author[0000-0003-2536-1614]{A. Calabr\'o}
\affil{INAF - OAR, via Frascati 33, 00078 Monte Porzio Catone, Roma, Italy} 
\author[0000-0003-1354-4296]{M. Llerena}
\affil{INAF - OAR, via Frascati 33, 00078 Monte Porzio Catone, Roma, Italy}

\begin{abstract}
The LyC escape fraction from galaxies, $\fesc$, is strongly boosted by galactic outflows. In the Attenuation-Free Model (AFM) accounting for the properties of $z>10$ galaxies, radiation-driven outflows develop once the galaxy specific star formation rate, $\rm sSFR \ge \sSFR^* \simeq 25\ \rm Gyr^{-1}$. As the cosmic sSFR increases with redshift, so does $\fesc(z)$, which, when globally averaged, grows from 0.007 to 0.6 in $0 < z < 20$. We successfully tested the model on specific data sub-samples.  Our predictions are consistent with measurements of $\fesc$  {up to $z=9.5$}, and provide a physical explanation for the observed decreasing trend of the mean UV galaxy spectral slope, $\beta$, towards high-$z$.     
\end{abstract}

\section{Introduction} \label{sec:intro}
Cosmic reionization is a key ingredient of structure formation theories encapsulating information on the nature and evolution of early light sources \citep[][]{Meiksin09, Dayal18, Gnedin22}. It also feeds back on galaxy formation by setting the filtering scale of baryonic perturbations \citep[][]{Gnedin00} below which gas is photo-evaporated from dark matter halos.

The rate at which Lyman continuum (LyC, $E > 13.6\ \rm eV$) ionizing photons are emitted into the intergalactic medium (IGM) at redshift $z$ is the product of the UV luminosity density, $\rho_{\rm UV}$, the ionizing photon production efficiency, $\xi_{\rm ion}$, and the fraction of ionizing photons escaping from the sources: $\dot n_{\rm ion}(z) = \fesc \xi_{\rm ion} \rho_{\rm UV}$. Generally, these factors depend on redshift and galaxy properties.

Breakthrough observations with the \textit{James Webb Space Telescope} (JWST) have significantly constrained\footnote{We note that some uncertainties on the UV luminosity function persist at $z > 10$ \citep[e.g.][]{Ferrara23, Ferrara24a}, where an unexpected overabundance of bright galaxies has been observed \citep[for a review, see][]{Adamo24}.} the evolution of both $\xi_{\rm ion}$ \citep[e.g.][]{Chen24, LLerena24, Simmonds24, Begley25, Pahl25} and $\rho_{\rm UV}$ \citep[e.g.][]{Bouwens22b,  Adams23, Donnan23a, Harikane23, McLeod23, Chemerynska24, Finkelstein24, Willott24}. In contrast, progress in determining $\fesc$ has been slower, owing to the intrinsic challenges of measuring this quantity due to LyC absorption by the neutral IGM  \citep[][]{Inoue14} during the Epoch of Reionization (EoR).

A combination of different physical processes, such as dust attenuation, ionization, gas density, and galaxy morphology are likely to control the value of $\fesc$ \citep[][]{Jaskot24, Mascia24}. However, a consensus has emerged that the presence of ionized, dust-transparent swaths in the galaxy is a necessary condition to achieve high LyC leakage. Such low-opacity channels are naturally created by outflows.
This scenario is supported by a number of observational \citep{Hogarth20, Vanzella22, Mainali22, Amorin24, Leclercq24}
and theoretical \citep[e.g.][]{decataldo:2017, Kimm19, Komarova21, Flury23} works.  

In short, it is believed that LyC escape occurs as a result of a two-stage process \citep{Hogarth20, Enders23, Flury24a} in which supernova (SN) explosions drive outflows carving low-density channels, that are more easily ionized by hard radiation from young stars, allowing their LyC to escape. 

However, it turns out that SN feedback might not be strictly necessary, and the process reduced to a single step. First of all, in a burst, the production of ionizing photons starts immediately, while SN feedback can be delayed up to $\approx 10$ Myr \citep[][]{Sukhbolt16, Jecmen23} depending on the upper mass limit for SNe which is not well constrained. Thus, as the LyC production drops dramatically after $3-4$ Myr, the opening of the SN channels might come too late. Moreover, catastrophic cooling may partially or completely quench SN feedback in high density environments as those of high-$z$ galaxies \citep[][]{Terlevich1992, Pizzati20, Ferrara24a}.   {Thus, an explanation purely based on the two-stage process might face some difficulties}. 

Fortunately, though, a simpler solution exists. Rather than by SNe, the ionized channels can be opened by outflows driven by radiation pressure from the same young stars producing the LyC photons, as pointed out also by \citet{Komarova21}, and recently confirmed by \citet{Carr25} from an analysis of the Low-z LyC Survey (LzLCS). This solution presents several advantages: (a) LyC production, photoionization and outflows start and evolve synchronously; (b) the outflow is momentum-driven by radiation, thus overcoming catastrophic cooling associated with energy-driven flows; (c) the outflow is powered by the much more numerous\footnote{The production rate of photons at 1500 \AA\ is $\approx 15 (10^{25} /\xi_{\rm ion} \rm [Hz\ erg^{-1}])$ times larger than the LyC one.} \textit{non-ionizing} UV photons absorbed by dust. We note that Ly$\alpha$ radiation pressure \citep[][]{Smith17, Kimm18, Tomaselli21, Nebrin25} acting directly on \HI might also contribute significantly to the driving.

As gas and dust are displaced on scales larger than the galaxy size by the radiation-driven outflow, the associated UV and LyC optical depth decrease makes the galaxy brighter and bluer. This is the pivotal idea behind the \quotes{Attenuation-Free Model} \citep[AFM,][]{Ferrara23, Ziparo23, Fiore23, Ferrara24a, Ferrara24b, Ferrara25a}. 

AFM consistently explains the overabundance of super-early ($z>10$) galaxies \citep{Ferrara23}, the evolution of the cosmic SFR density \citep{Ferrara24a}, the star formation history and properties of individual galaxies \citep{Ferrara24b}, the occurrence of mini-quenched galaxies, \citep{Gelli23}. It also makes specific predictions for \textit{Atacama Large Millimeter Array} (ALMA) targeted observations of super-early sources \citep{Ferrara25a}.  
Here, we extend the predictive power of AFM to derive the redshift evolution of LyC escape fraction, $\fesc(z)$, from galaxies. 

\section{Model} \label{sec:model}
The basic idea behind AFM is that galaxies can go through evolutionary phases in which their bolometric luminosity, $L_{\rm bol}$,  becomes super-Eddington,
\begin{equation}
L_{\rm bol} > L_E^{\rm eff} = A^{-1} L_E, 
\label{eq:superEdd}
\end{equation}
where $L_{E}= (4\pi G m_p c /\sigma_T) M_* = 1.26\times 10^{38} (M_*/M_\odot)\, \rm erg\, s^{-1}$ is the classical Eddington luminosity. This is reduced by a \quotes{boost} factor $A \approx \sigma_d/\sigma_T$, i.e. the ratio of the dust-to-Thomson cross-section in a dusty medium. \citet{Ferrara24a} and \citet{Nakazato25} find that $A = 100-1000$, mainly depending on gas metallicity and column density. 

As $L_{\rm bol}$ is proportional to the star formation rate (SFR) and $L_E\propto M_*$, \citet{Ferrara24a} showed that the super-Eddington condition eq. \ref{eq:superEdd}  translates into one on the specific star formation rate, ${\rm sSFR=SFR}/M_\star$: 
\begin{equation}
{\rm sSFR} > {\rm sSFR}^\star \simeq 25 \left(\frac{100}{A}\right) \left(\frac{2}{f_{\rm bol}}\right) {\rm Gyr^{-1}},
\label{eq:ssfr_thresh}
\end{equation}
where $f_{\rm bol}$ is the bolometric to UV correction. 
Eq. \ref{eq:ssfr_thresh} represents the necessary condition for a galaxy to develop a radiation-driven outflow ejecting both dust and gas from the system.

In AFM (see \citealt{Ferrara24a}) the specific star formation rate is simply written as  
\be
{\rm sSFR} = 0.64\, \frac{\epsilon_\star}{\langle\epsilon_\star\rangle}\,(1+z)^{3/2}\, \rm Gyr^{-1},
\label{eq:sSFR}
\ee
where $\epsilon_\star$ is the instantaneous SF efficiency, $\langle\epsilon_\star\rangle$ is the efficiency averaged over a free-fall time.
Note that if $\epsilon_\star \simeq \langle\epsilon_\star\rangle$, the sSFR is a function of redshift only, and grows monotonically.  Combining eqs. \ref{eq:ssfr_thresh} and \ref{eq:sSFR} leads to conclude that \textit{on average} galaxies become super-Eddington for $z \simgt 10$. 

  {Experimental evidence that the fraction of super-Eddington galaxies increases with redshift is brought by \citet{Boyett24}. These authors identified Extreme Emission Line Galaxies (EELG, with optical rest-frame equivalent width lines EW~$> 750$ \AA) from a sample of 85 galaxies in $3 < z < 9.5$. They find that the fraction of galaxies in the EELG phase increases from 23\% at $3 < z <4.1$ to  61\% at $z>5.7$. As these authors also show (see their Fig. 13) that virtually all EELG galaxies have sSFR $> 25\ \rm Gyr^{-1}$, we can conclude that the fraction of super-Eddington galaxies increases with redshift, as found here.}

Eq. \ref{eq:sSFR} provides a very good match to the mean observed sSFR trend (see Fig. 2 in \citealt{Ferrara24a}), but it does not account for the observed sSFR scatter at fixed redshift. The scatter naturally arises in the model if the efficiency varies over $t_{\rm ff}$, i.e. $\langle \epsilon_\star \rangle \neq \epsilon_\star$, as expected for stochastic star formation histories \citep{Pallottini23, Shen23, Mirocha23, Cole25}, or multiple subsequent phases of quiescent and active star formation \citep{Kobayashi23, Gelli25, Looser23, Baker25}.  

We take an empirical approach, and adopt the scatter of the observed sSFR in the well-studied range $8 < z < 10$, which corresponds to a fractional standard deviation of 83\% (see discussion in \citealt{Ferrara24a}). Further assuming that the sSFR is normally distributed, we compute the fraction of super-Eddington galaxies, as
\begin{equation}
f_E(z \vert {\rm sSFR}^*) = \frac{1}{2} {\rm erfc} \left(\frac{{\rm sSFR}^* - {\rm sSFR}(z)}{\sqrt{2}\sigma}\right)
    \label{eq:fE}
\end{equation}
at any given redshift. We find that $f_{E}$ increases with redshift. For example, at  $z=6\ (14)$ we find $f_{E}=0.22\ (0.76)$ for the fiducial value $\sSFR^* = 25\ \rm Gyr^{-1}$. 

%
%
%
%
\begin{figure*}
\centering\includegraphics[width = 1.0 \textwidth]{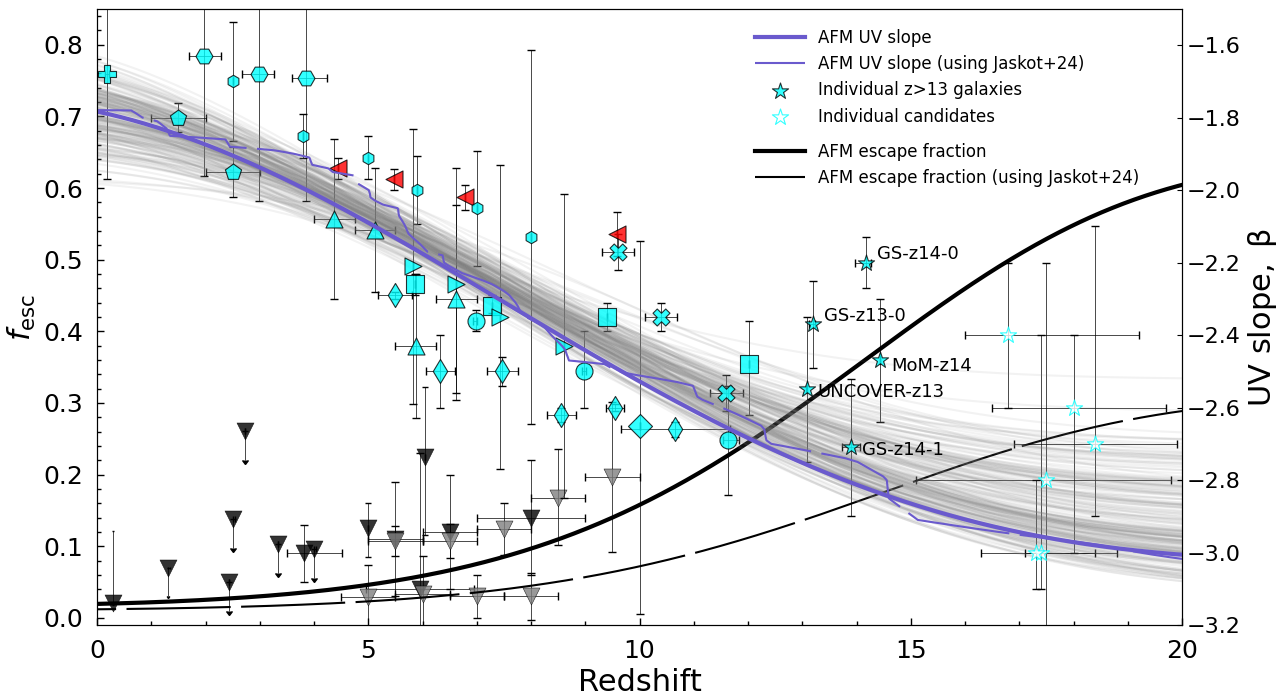}
\caption{  
AFM model predictions for the redshift evolution of $\fesc$ (solid black line) and $\beta$ (solid blue) adopting the \citet{Chisholm22} relation eq. \ref{eq:Chisholm}. The grey lines show 200 random samples drawn from the MCMC chains for this case. The thin dashed black and blue lines show the same predictions using the \citet[][implemented in \citealt{Flury24}]{Jaskot24} relation. Predictions are compared with  {available $f_{\rm esc}$ data (black downward triangles)} obtained from stacked (or averaged) direct \citep[][]{Grazian17, Marchi17, Alavi20, Flury22, Wang25}, indirect \citep[][]{Mascia24, Mascia25}, and QSO fields \citep[][]{Cain25}  measurements  {(see footnote \ref{fn:methods} for a brief description)} of $\fesc$ in $2 < z < 9$. 
Gray points are SED fitting-based $\fesc$ data (\citealt{Papovich25}; Giovinazzo et al. in prep; we have increased the statistical errors by $\approx 3\times$  to account for systematic uncertainties deriving from different SFH assumptions or codes in the SED fitting, see \citealt{Helton24}). We also show the comparison with UV slope data (cyan/red points) from \citet[][plus]{Overzier11}, \citet[][small hexagon]{Bouwens14}, \citet[][pentagon]{Kurczynski14}, \citet[][hexagon]{Morales24}, \citet[][triangle up]{Nanayakkara23}, \citet[][circle]{Austin24}, \citet[][square]{Topping23}, \citet[][thin diamond]{Roberts24}, \citet[][triangle right]{Saxena24},  \citet[][cross]{Cullen24}, \citet[][diamond]{Morales24b}. The \citet[][triangle left]{Dottorini24} dataset, separately discussed in Sec. \ref{subsec:dottorini}, is highlighted in red. For reference, we have reported $\beta$ values of spectroscopically confirmed $z>13$ individual galaxies (cyan stars) that are not part of the other samples. These are taken from \citet[][UNCOVER-z13]{Fujimoto23b}; \citet[][GS-z13-0]{Curtis23}; \citet[][GS-z14-0, GS-z14-1]{Carniani24a}; Oesch, priv. comm. (MoM-z14, $z=14.44$). Finally, $z\approx 17$ photometric candidates by \citet[][]{Perez25} are shown by open stars. 
}
\label{fig:fesc_vs_z}
\end{figure*}

\subsection{A physical functional form for the escape fraction}
Armed with $f_E(z)$ given by eq. \ref{eq:fE}, we propose a heuristic model for the LyC escape fraction whose physical basis is clarified by an example. 
\citet{Topping_2022} searched for galaxies with very blue UV colors in \textit{JWST}/NIRCam imaging of the EGS field.  Three systems with stellar mass $\approx 5 \times 10^7\ M_\odot$ exhibit both extremely blue UV slopes $\beta \approx -3$, and weak nebular line emission that can only be explained with large values of $\fesc = 0.6-0.8$. These systems are also very compact ($r_e \simlt 250$ pc). Noticeably, all the three systems have very large $\sSFR = (47,99,239)\ \rm Gyr^{-1}$, i.e. well above the AFM threshold value sSFR$^* \simeq 25\ \rm Gyr^{-1}$. 

A natural interpretation is that these sources are experiencing a super-Eddington phase, during which both the dust and the gas are displaced by an outflow to larger scales, thus producing the blue spectrum, and favoring LyC escape. 
Inspired by this evidence, we postulate that at each redshift the value of $\fesc(z)$ is the weighted mean performed on super- and sub-Eddington populations:
\begin{equation}
\fesc(z) = [1-f_E(z \vert {\rm sSFR}^*)] f_0 + f_E(z \vert {\rm sSFR}^*) f_1,
\label{eq:fesc}
\end{equation}
where $f_0$ and $f_1$ (with $f_1 \gg f_0$) are the escape fractions of sub- and super-Eddington galaxies, respectively. 

Interestingly, a bimodal $\fesc$ distribution has also been found in simulations \citep[e.g.][]{Kostyuk25}, albeit possibly of different physical nature. Bimodality has also been often invoked observationally to explain why a significant amount of LyC flux escapes from some galaxies, while the dominant population shows no significant leakage \citep{Grazian17, Nakajima20}. However, we remark that no sign of bimodality is evident in the $\fesc$ distribution from  the LzLCS sample \citep[][]{Flury22}.

\subsection{Fixing the free parameters}
To fix the free parameters we exploit the large amount of high-quality measurements of the UV slope, $\beta$, for galaxies in the redshift range $0 < z < 12$, most of which have become available thanks to \textit{JWST} observations.  {These data (cyan points), along with stacked direct measurements or indirect inference determinations based on multi-variate diagnostics (black)} and SED fitting-based (gray) measurements of $\fesc$ are shown in Fig. \ref{fig:fesc_vs_z}.

We determine $f_0$, $f_1$ and sSFR$^*$ by fitting the model eq. \ref{eq:fE} to the UV slope data using the affine-invariant ensemble sampler for Markov Chain Monte Carlo (MCMC) proposed by \citet{Goodman10} and implemented in python as the open source code \code{emcee} by \citet{EMCEE13}. To enable such procedure, we transform the predicted $\fesc$ into a $\beta$ value by inverting the  \citet{Chisholm22} relation,
\begin{equation}\label{eq:Chisholm}
\fesc = (1.3 \pm 0.6) \times 10^{-4} \times 10^{-(1.22 \pm 0.1)\beta},    
\end{equation}
and propagate the uncertainties into the final posterior distribution by sampling them during the MCMC procedure. 
 {We caution the reader that the relation is strictly valid for $\beta \simgt -2.6$, and we extrapolate it to bluer slopes.} 
  
\section{Results and tests} \label{sec:results}
The MCMC fitting procedure provides the following best-fit values (16$^{\rm th}$, 50$^{\rm th}$, and 84$^{\rm th}$ percentiles) of the three free parameters:
$f_0 = 0.007_{-0.005}^{+0.008};\ f_1 = 0.643_{-0.207}^{+0.224};\ {\sSFR}^* = 35.439_{-4.352}^{+3.313}\ \rm Gyr^{-1}$. We have assigned flat priors to the parameters, i.e. $0< f_0 (f_1) < 1$, $0 < \sSFR^*\ \rm [Gyr^{-1}] < 100$. 
The above result already highlights some key points of our study. 

First, the presence or absence of outflows identifies two classes of galaxies: strong LyC leakers with $\fesc = f_1 \simgt 40\%$, and LyC-dark ones ($\fesc = f_0 \simlt 1\%$). Their variable redshift mixture, governed by the evolution of the sSFR, yields the cosmic global $\fesc(z)$ value.

Secondly, although AFM physical arguments indicate a fiducial value of sSFR$^* \simeq 25\ \rm Gyr^{-1}$, this parameter can vary in the (prior) range $1-100\ \rm Gyr^{-1}$. Yet, the recovered value is very close, within uncertainties on $A$ discussed in \citet{Nakazato25}, to the expected one, lending support to the AFM physical framework.

The best-fit values $(f_0, f_1, \sSFR^*)$, once inserted in eq. \ref{eq:fesc}, yield the AFM redshift evolution of $\fesc$ (Fig. \ref{fig:fesc_vs_z}, black line). The escape fraction grows gradually towards high redshift, driven by the cosmic increase of sSFR in galaxies, and reaches values close to 60\% at $z=20$. This prediction is in excellent agreement with direct, indirect, and SED fitting-based measurements{\footnote{{Direct methods measure $\fesc$ from the LyC flux; indirect methods use a combination of indicators (such as the $\beta$ slope, EW(H$\beta$), O32); QSO-field methods obtain $\fesc$ from the ionising background and LyC photon mean free path from quasar spectra; SED-fitting methods include $\fesc$ as one of the free parameters to reproduce the galaxy photometry/spectrum.}\label{fn:methods}}}. 

In addition to $\fesc$, AFM can predict also the redshift evolution of the UV slope, which is compared with a large set of data in Fig. \ref{fig:fesc_vs_z}. The best-fitting curve, deduced from the predicted $\fesc(z)$ in combination with the Chisholm relation eq. \ref{eq:Chisholm}, matches very nicely the data. It also provides a physical interpretation of the observed $\beta(z)$ trend: as outflows become more frequent at early epochs (recall that $f_E$ increases with $z$), more galaxies manage to efficiently disperse the gas and dust trapping LyC photons, at the same time showing bluer colors. Thus, we conclude that outflows play a key role in governing the UV slope evolution.

For reference we have also plotted in Fig. 1 the $\beta$ values of spectroscopically confirmed $z > 13$ individual galaxies that are not included in lower-$z$ samples, along with the recently discovered $z\approx 17$ candidates \citep[][]{Perez25}. 

Some of these galaxies tend to have slightly redder colors than the model, representing a global mean. However, we recall that we made no attempts to correct for the generally small reddening potentially introduced by the presence of nebular continuum in these galaxies \citep[][]{Topping23}. As $\beta$ gets closer to $-2.5$, nebular continuum, rather than dust, becomes the dominant reddening source \citep[][]{Narayanan24}  Moreover, we stress that the Chisholm relation has been validated only for $\beta \simlt -2.6$. 

In any case, it is tantalizing to see that the trend of decreasing $\beta$ with increasing $\fesc$ is retained passing from spectroscopically confirmed galaxies at $z=13-14$ ($\fesc\approx 30$\%) to candidates at $z=17-18$ where $\fesc\approx 50\%$.

We have repeated the same procedure using the $\beta - \fesc$ relation obtained by \citet{Jaskot24}. Such relation does not have a simple analytical expression. It closely resembles eq. \ref{eq:Chisholm} for $\beta \simgt -2$, but predicts a lower $\fesc$ for bluer $\beta$ values. Using such relation, we obtain a very good fit to the UV slope data with    $f_0 = 0.007_{-0.004}^{+0.006};\ f_1 = 0.332_{-0.148}^{+0.301};\ {\sSFR}^* = 36.688_{-6.920}^{+6.957}\ \rm Gyr^{-1}$. The most noticeable difference with respect to \cite{Chisholm22} is the low value of $f_1$, implying a 2$\times$ lower $\fesc$ for super-Eddington galaxies.  {In spite of the excellent $\beta(z)$ match to the data (see Fig. \ref{fig:fesc_vs_z}), the \citet{Jaskot24} relation somewhat underestimates (see dashed black line) the $\fesc$ data, although reproducing a similar trend. For this reason, we adopt eq. \ref{eq:Chisholm} as our fiducial relation, although with current data we cannot solidly discriminate between the two. } 

%
%
%
%
\begin{figure*}
\centering
\begin{minipage}{0.47\textwidth}
    \centering
    \includegraphics[width=\textwidth]{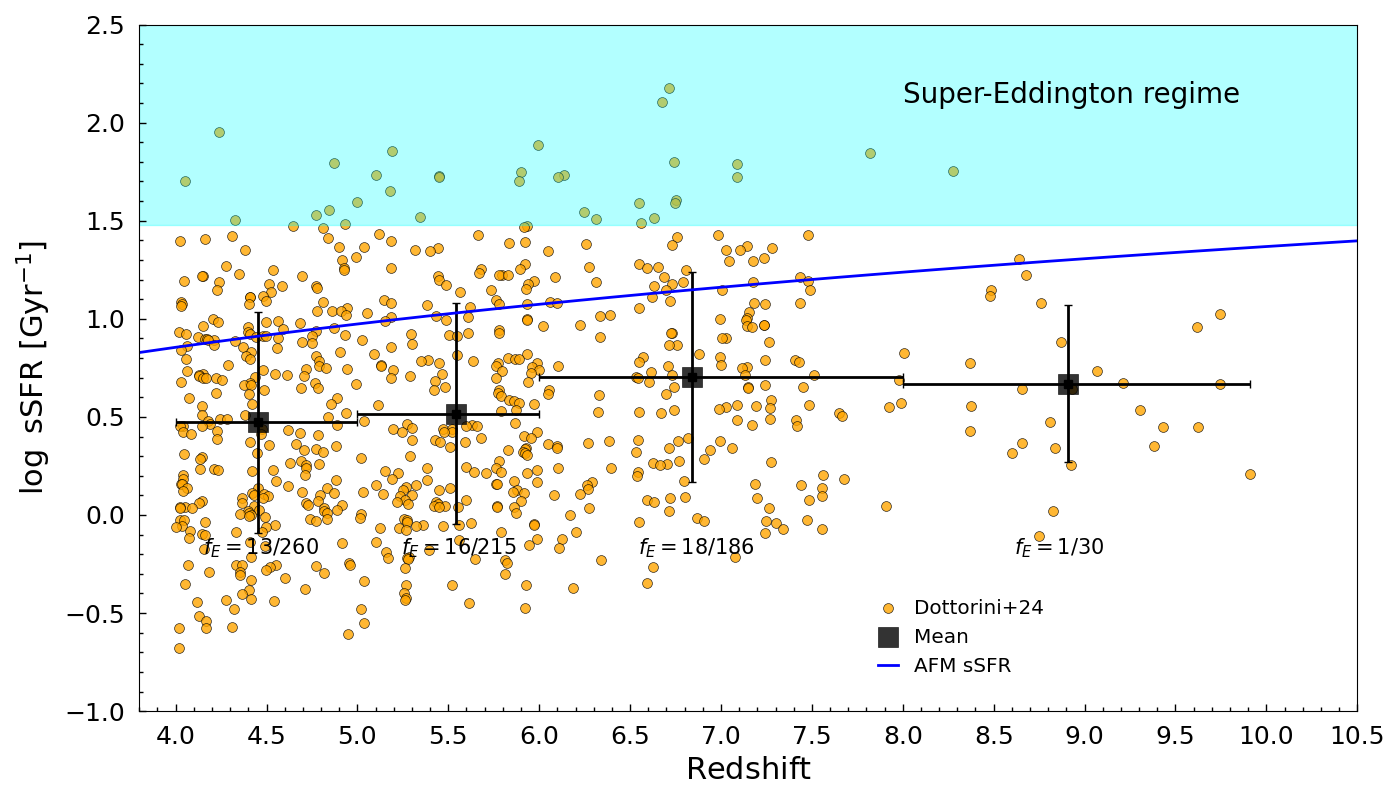}
    \caption{
    Specific star formation rate of the D24 sample (691 galaxies, yellow points); the black points show the mean sSFR in four redshift bins. Also shown is the AFM sSFR evolution (eq. \ref{eq:sSFR}), and the Super-Eddington regime region.
    }
    \label{fig:dottorini_ssfr}
\end{minipage}%
\hfill
\begin{minipage}{0.47\textwidth}
    \centering
    \includegraphics[width=\textwidth]{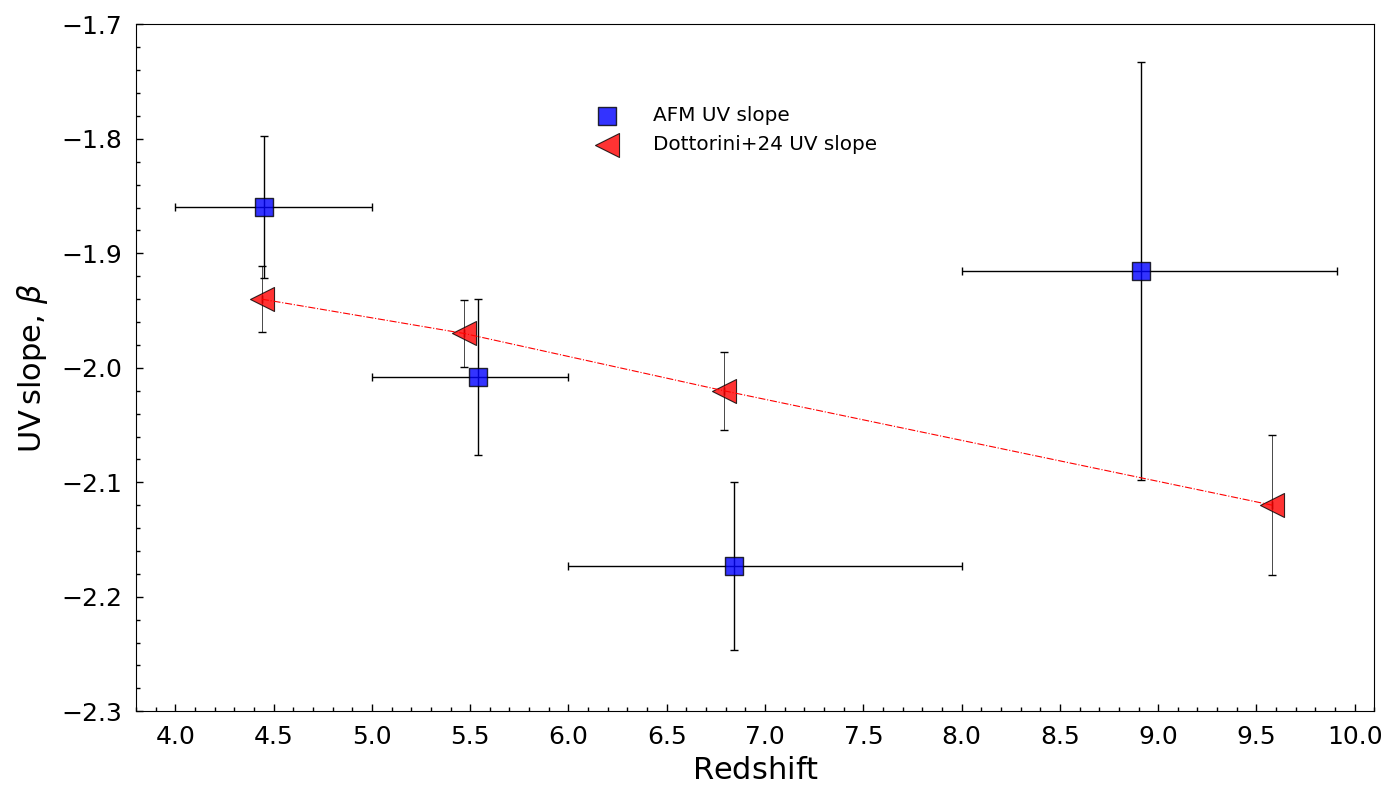}
    \caption{
    Redshift evolution of the UV slope, $\beta$, for the D24 sample (red points) compared to the analogous one predicted by the attenuation-free model (blue) using the super-Eddington fraction, $f_E^{D24}$, of the D24 sample (see text).}
    \label{fig:dottorini_beta}
\end{minipage}
\end{figure*}

\subsection{Test on specific data samples}\label{subsec:dottorini}
From Fig. \ref{fig:fesc_vs_z} we see that the \citet[][D24]{Dottorini24} data are on average redder ($\beta\approx -2$) than the rest of the data at the same redshift. It is then instructive to analyse the origin of such discrepancy and use it as a sanity check of our model. 

Fig. \ref{fig:dottorini_ssfr} shows the sSFR (averaged over the last $1$ Myr) of the 691 galaxies in the D24 sample as a function of redshift, along with the AFM prediction eq. \ref{eq:sSFR} (blue curve). The D24 sample is selected spectroscopically, and therefore complete only down to $M_{\rm UV} =-18.5$ at $z=8.5$.
We group the galaxies in 4 bins of mean redshift $\langle z \rangle = (4.45, 5.54, 6.84, 8.91)$, and compute the corresponding mean $\log \rm sSFR [Gyr^{-1}] = (0.47, 0.52, 0.70, 0.67)$. These values are lower by $\approx 0.3-0.5$ dex than expected from eq. \ref{eq:sSFR}, which, however, matches the overall observed sSFR evolution. 

Most importantly, though, the fraction of super-Eddington galaxies, also shown in Fig. \ref{fig:dottorini_ssfr} in the four D24 bins, is $f_E^{D24} = (5, 7, 10, 3)\times 10^{-2}$, i.e. about 2 to 8 (in the highest-$z$ bin) times smaller than predicted by AFM (eq. \ref{eq:superEdd}), thus explaining why the D24 sample is characterized by redder UV slopes. 

To quantify this statement, in Fig. \ref{fig:dottorini_beta} we compare AFM predictions and D24 measurements of $\beta(z)$. In practice, we have inserted $f_E^{D24}$ in eq. \ref{eq:fesc} to derive $\fesc$, and combined it with eq. \ref{eq:Chisholm}. We find a remarkable agreement between theory and observations, given the uncertainties involved. This test demonstrates AFM's predictive power.

\section{Discussion and implications} \label{sec:discuss}
The above results confirm the key role of outflows in creating ionized, dust-transparent channels through which LyC photons can leak. In turn, radiation-driven outflows preferentially develop in compact galaxies with high sSFR. 
These conditions are increasingly met towards high-redshift, due to the overall galaxy size decrease and increasing ${\rm sSFR} \propto (1+z)^\alpha$, with $\alpha \approx 1.5$. 
A simple model based on this idea, implemented in eq. \ref{eq:fesc}, fits exceptionally well the observed redshift evolution of $\fesc$ and UV slope of galaxies.

Prototypical examples of the relation between $\fesc$ and sSFR are the three galaxies (EGS-42501, EGS-41419, and EGS-47688) observed with {JWST}/NIRCam in the redshift range $7.2 < z < 8.4$ \citep[][ {method: SED-fitting}]{topping2022}. These exhibit extremely blue UV slopes ($\beta \approx -3.1$), very compact ($r_e <260$ pc) sizes, and very large LyC escape fractions ($f_{\rm esc}=0.8, 0.57, 0.65$). Remarkably, their sSFR is significantly super-Eddington, $\rm sSFR = (47, 99, 239)\ Gyr^{-1}$. Another outstanding case is EBG-1, a spectroscopically confirmed galaxy at $z = 9.25$ with a very blue slope, $\beta =-2.99 \pm 0.15$, and a super-Eddington $\rm sSFR$ of $57\ \rm Gyr^{-1}$ for which \citet[][indirect]{Yanagisawa24} have inferred $\fesc \simgt 0.5$.

Although statistically more rare, systems with large sSFR showing LyC leakage can be found also at lower redshifts. For example, \citet[][  direct, see their Fig. 17]{Flury22} using the Low-redshift Lyman Continuum Survey (LzLCS) find that no $z \approx 0.3$ LyC emitters with $f_{\rm esc}> 10\%$ have $\rm sSFR < 10\ Gyr^{-1}$. 

 {Additional examples at $z \approx 0.4$ are provided by \citet[][  direct]{Izotov21} who report nine strong LyC leakers with sSFR of $150-630\ \rm Gyr^{-1}$. These authors also favour radiation pressure as the most likely explanation for the high $\fesc$ observed \citep[][]{Izotov18}. Also notable is the galaxy J1154+2443 at $z = 0.37$ with $\fesc=0.46$, and a super-Eddington $\rm sSFR = 120\ \rm Gyr^{-1}$ \citep[][  direct]{Izotov18b}.}

There is also a growing evidence for extreme intermediate-$z$ LyC emitters. \citet[][  direct]{Marques22, Marques24} reports the case of  J1316+2614 at $z = 3.613$, a compact, UV-bright, super-Eddington galaxy with $\rm sSFR \approx 100\, Gyr^{-1}$. The UV slope is very steep ($\beta \approx -2.6$) and the inferred $\fesc = 0.89$. Another striking example is the Sunburst Arc \citep[][  direct]{Rivera19} a single, compact galaxy at redshift $z = 2.37$ lensed into 12 images, each of them with $\fesc$ in the range $0.19-0.64$. At $z = 3.2$, \citet[][  direct]{Vanzella16} 
detected a compact Green Pea galaxy (Ion2) that shows an $\fesc \simgt 0.5$; this discovery was supplemented with the discovery of Ion3 at $z=4$, with a measured $\fesc \simgt 0.6$ \citep[][  direct]{Mestric25}.
Both these sources are very blue, compact and show a high sSFR.

We conclude that available data seem to support the idea that compactness, blue colors and high sSFR, according to AFM, lead to the onset of outflows clearing the path to LyC photons, eventually resulting in high $\fesc$ values. However, we warn that AFM cannot predict the $\fesc$ of individual sources as it is intrinsically based on globally averaged quantities, e.g. the evolution of the cosmic sSFR in eq. \ref{eq:sSFR}. 

Several galaxy properties, such as metallicity, dust content or HI column densities, are likely to introduce scatter in $\fesc$. Geometry might also play a role. For example the channels carved by the outflows might be anisotropic and cover a limited solid angle \citep[][]{Flury24a, Ji25}. This could explain why the best-fit escape fraction for super-Eddington galaxies we find is $f_1 \approx 44-87 \%$ rather than a 100\% leakage expected from a spherical outflow geometry. 

The implications of the $\fesc(z)$ evolution obtained here for reionization must be explored in detail. A solid conclusion, though, is that $\fesc$ evolves significantly with redshift and cannot be simply assumed to be constant. 

We also point out that among the $\approx 20$ spectroscopically confirmed galaxies at $z>10$ (collected in \citealt{Ferrara25b},  {and augmented with the two newly discovered CAPERS sources \citealt{Kokorev25}}) only two of them (GN-z-11 and GHZ2) show prominent UV emission lines. This fact is surprising given the large SFR and blue colors of these
sources. A possible explanation is LyC photons responsible for UV lines excitation partly or largely escape from the bright super-early galaxies observed so far \citep[][]{Hainline24}. Alternatively, or in addition, these systems might have low metallicity \citep{Curti24}, or be in a post-starburst phase in which the outflow has already ceased \citep{Ferrara24b, Endsley24}.  

Finally, a large $\fesc$ would result in larger ionized IGM bubbles around these sources, which in turn might explain the surprising Ly$\alpha$ line detections from sources such as GN-z-11 ($z=10.6$, \citealt{Bunker23}) and GS-z13-1-LA ($z=13.0$, \citealt{Witstok23})


\section{Summary} \label{sec:summary}
We have used the Attenuation-Free Model to predict the redshift evolution of the LyC escape fraction from galaxies.
The basic idea is that LyC leakage can occur predominantly through ionized, dust-transparent channels carved by outflows. AFM predicts that outflows powered by radiation pressure from young stars in compact galaxies when the sSFR exceeds the super-Eddington threshold $\rm sSFR^* \simeq 25\ Gyr^{-1}$. We then postulate that at each redshift the value of $\fesc(z)$ is the weighted mean performed over super- and sub-Eddington populations. We require that such model, combined with the most recent $\fesc$-UV slope relations, fits simultaneously the observed redshift evolution of these two quantities. With this procedure we find that:
\begin{itemize}
\item[{\color{red} $\blacksquare$}] The presence or absence of outflows identifies two classes of galaxies: strong LyC leakers with $\fesc = f_1 \simgt 40\%$, and LyC-dark ones ($\fesc = f_0 \simlt 1\%$). Their variable redshift mixture, governed by the evolution of the sSFR, yields the cosmic global $\fesc(z)$ value.

\item[{\color{red} $\blacksquare$}] The escape fraction grows gradually towards high redshift, driven by the cosmic increase of sSFR in galaxies, and reaches values close to 60\% at $z=20$. This prediction is in excellent agreement with direct, indirect, and SED fitting-based measurements of $\fesc$ in $2 < z < 9$.

\item[{\color{red} $\blacksquare$}] The model, in combination with the Chisholm relation eq. \ref{eq:Chisholm}, provides an excellent match to the UV-slope data from $z=0$ to $z=12$. It also provides a physical interpretation of the observed $\beta(z)$ trend: as outflows become more frequent at early epochs more galaxies manage to efficiently disperse the gas and the dust trapping LyC photons, at the same time showing bluer colors.  {Thus, we conclude that outflows, among other relevant properties such as metallicity, dust
content or HI column densities, play a key role in governing the UV slope evolution.}

\item[{\color{red} $\blacksquare$}] We have performed successful tests on specific samples (e.g. \citealt{Dottorini24}) showing red-ward deviations from the observed $\beta(z)$ trend. We find that the deviation is due to the lower-than-average sSFR of the sample, as expected from AFM.

\item[{\color{red} $\blacksquare$}] The implications of the $\fesc(z)$ evolution obtained here for reionization must be explored in detail. A solid conclusion, though, is that $\fesc$ evolves significantly, and cannot be simply assumed to be redshift-independent. Large $\fesc$ values deep in the EoR might explain the weaker or absent UV emission lines and Ly$\alpha$ detections from super-early galaxies. They also favour an early start of the reionization process \citep[see e.g.][]{Munoz24}. 

\end{itemize}

\acknowledgments
We thank S. Flury, S. Fujimoto, E. Giovinazzo, S. Mascia, P. Oesch  for providing key experimental data, and S. Carniani, V. d'Odorico for useful comments. This work is supported by the ERC Advanced Grant INTERSTELLAR H2020/740120, and in part by grant NSF PHY-2309135 to the Kavli Institute for Theoretical Physics. 
This work made use of the MCMC sampler EMCEE \citep[][]{EMCEE13}.
Plots are produced with the \textsc{matplotlib} \citep{Hunter07} package.

\bibliographystyle{aasjournal}
\bibliography{paper}

\begin{thebibliography}{}
\expandafter\ifx\csname natexlab\endcsname\relax\def\natexlab#1{#1}\fi
\providecommand{\url}[1]{\href{#1}{#1}}
\providecommand{\dodoi}[1]{doi:~\href{http://doi.org/#1}{\nolinkurl{#1}}}
\providecommand{\doeprint}[1]{\href{http://ascl.net/#1}{\nolinkurl{http://ascl.net/#1}}}
\providecommand{\doarXiv}[1]{\href{https://arxiv.org/abs/#1}{\nolinkurl{https://arxiv.org/abs/#1}}}

\bibitem[{{Adamo} {et~al.}(2024){Adamo}, {Atek}, {Bagley}, {Ba{\~n}ados}, {Barrow}, {Berg}, {Bezanson}, {Brada{\v{c}}}, {Brammer}, {Carnall}, {Chisholm}, {Coe}, {Dayal}, {Eisenstein}, {Eldridge}, {Ferrara}, {Fujimoto}, {de Graaff}, {Habouzit}, {Hutchison}, {Kartaltepe}, {Kassin}, {Kriek}, {Labb{\'e}}, {Maiolino}, {Marques-Chaves}, {Maseda}, {Mason}, {Matthee}, {McQuinn}, {Meynet}, {Naidu}, {Oesch}, {Pentericci}, {P{\'e}rez-Gonz{\'a}lez}, {Rigby}, {Roberts-Borsani}, {Schaerer}, {Shapley}, {Stark}, {Stiavelli}, {Strom}, {Vanzella}, {Wang}, {Wilkins}, {Williams}, {Willott}, {Wylezalek}, \& {Nota}}]{Adamo24}
{Adamo}, A., {Atek}, H., {Bagley}, M.~B., {et~al.} 2024, arXiv e-prints, arXiv:2405.21054, \dodoi{10.48550/arXiv.2405.21054}

\bibitem[{{Adams} {et~al.}(2023){Adams}, {Conselice}, {Ferreira}, {Austin}, {Trussler}, {Juod{\v{z}}balis}, {Wilkins}, {Caruana}, {Dayal}, {Verma}, \& {Vijayan}}]{Adams23}
{Adams}, N.~J., {Conselice}, C.~J., {Ferreira}, L., {et~al.} 2023, \mnras, 518, 4755, \dodoi{10.1093/mnras/stac3347}

\bibitem[{{Alavi} {et~al.}(2020){Alavi}, {Colbert}, {Teplitz}, {Siana}, {Scarlata}, {Rutkowski}, {Mehta}, {Henry}, {Dai}, {Haardt}, \& {Bagley}}]{Alavi20}
{Alavi}, A., {Colbert}, J., {Teplitz}, H.~I., {et~al.} 2020, \apj, 904, 59, \dodoi{10.3847/1538-4357/abbd43}

\bibitem[{{Amor{\'\i}n} {et~al.}(2024){Amor{\'\i}n}, {Rodr{\'\i}guez-Henr{\'\i}quez}, {Fern{\'a}ndez}, {V{\'\i}lchez}, {Marques-Chaves}, {Schaerer}, {Izotov}, {Firpo}, {Guseva}, {Jaskot}, {Komarova}, {Mu{\~n}oz-Vergara}, {Oey}, {Bait}, {Carr}, {Chisholm}, {Ferguson}, {Flury}, {Giavalisco}, {Hayes}, {Henry}, {Ji}, {King}, {Leclercq}, {{\"O}stlin}, {Pentericci}, {Saldana-Lopez}, {Thuan}, {Trebitsch}, {Wang}, {Worseck}, \& {Xu}}]{Amorin24}
{Amor{\'\i}n}, R.~O., {Rodr{\'\i}guez-Henr{\'\i}quez}, M., {Fern{\'a}ndez}, V., {et~al.} 2024, \aap, 682, L25, \dodoi{10.1051/0004-6361/202449175}

\bibitem[{{Austin} {et~al.}(2024){Austin}, {Conselice}, {Adams}, {Harvey}, {Duan}, {Trussler}, {Li}, {Juodzbalis}, {Ormerod}, {Ferreira}, {Westcott}, {Harris}, {Wilkins}, {Bhatawdekar}, {Caruana}, {Coe}, {Cohen}, {Driver}, {D'Silva}, {Frye}, {Furtak}, {Grogin}, {Hathi}, {Holwerda}, {Jansen}, {Koekemoer}, {Marshall}, {Nonino}, {Ortiz}, {Pirzkal}, {Robotham}, {Ryan}, {Summers}, {Willmer}, {Windhorst}, {Yan}, \& {Zackrisson}}]{Austin24}
{Austin}, D., {Conselice}, C.~J., {Adams}, N.~J., {et~al.} 2024, arXiv e-prints, arXiv:2404.10751, \dodoi{10.48550/arXiv.2404.10751}

\bibitem[{{Baker} {et~al.}(2025){Baker}, {D'Eugenio}, {Maiolino}, {Bunker}, {Simmonds}, {Tacchella}, {Witstok}, {Arribas}, {Carniani}, {Charlot}, {Chevallard}, {Curti}, {Curtis-Lake}, {Jones}, {Kumari}, {Rinaldi}, {Robertson}, {Williams}, {Willott}, \& {Zhu}}]{Baker25}
{Baker}, W.~M., {D'Eugenio}, F., {Maiolino}, R., {et~al.} 2025, arXiv e-prints, arXiv:2501.09070, \dodoi{10.48550/arXiv.2501.09070}

\bibitem[{{Begley} {et~al.}(2025){Begley}, {McLure}, {Cullen}, {McLeod}, {Dunlop}, {Carnall}, {Stanton}, {Shapley}, {Cochrane}, {Donnan}, {Ellis}, {Fontana}, {Grogin}, \& {Koekemoer}}]{Begley25}
{Begley}, R., {McLure}, R.~J., {Cullen}, F., {et~al.} 2025, \mnras, 537, 3245, \dodoi{10.1093/mnras/staf211}

\bibitem[{{Bouwens} {et~al.}(2014){Bouwens}, {Illingworth}, {Oesch}, {Labb{\'e}}, {van Dokkum}, {Trenti}, {Franx}, {Smit}, {Gonzalez}, \& {Magee}}]{Bouwens14}
{Bouwens}, R.~J., {Illingworth}, G.~D., {Oesch}, P.~A., {et~al.} 2014, \apj, 793, 115, \dodoi{10.1088/0004-637X/793/2/115}

\bibitem[{{Bouwens} {et~al.}(2022){Bouwens}, {Stefanon}, {Brammer}, {Oesch}, {Herard-Demanche}, {Illingworth}, {Matthee}, {Naidu}, {van Dokkum}, \& {van Leeuwen}}]{Bouwens22b}
{Bouwens}, R.~J., {Stefanon}, M., {Brammer}, G., {et~al.} 2022, arXiv e-prints, arXiv:2211.02607, \dodoi{10.48550/arXiv.2211.02607}

\bibitem[{{Boyett} {et~al.}(2024){Boyett}, {Bunker}, {Curtis-Lake}, {Chevallard}, {Cameron}, {Jones}, {Saxena}, {Charlot}, {Curti}, {Wallace}, {Arribas}, {Carniani}, {Willott}, {Alberts}, {Eisenstein}, {Hainline}, {Hausen}, {Johnson}, {Rieke}, {Robertson}, {Stark}, {Tacchella}, {Williams}, {Chen}, {Egami}, {Endsley}, {Kumari}, {Laseter}, {Looser}, {Maseda}, {Scholtz}, {Shivaei}, {Simmonds}, {Smit}, {{\"U}bler}, \& {Witstok}}]{Boyett24}
{Boyett}, K., {Bunker}, A.~J., {Curtis-Lake}, E., {et~al.} 2024, \mnras, 535, 1796, \dodoi{10.1093/mnras/stae2430}

\bibitem[{{Bunker} {et~al.}(2023){Bunker}, {Saxena}, {Cameron}, {Willott}, {Curtis-Lake}, {Jakobsen}, {Carniani}, {Smit}, {Maiolino}, {Witstok}, {Curti}, {D'Eugenio}, {Jones}, {Ferruit}, {Arribas}, {Charlot}, {Chevallard}, {Giardino}, {de Graaff}, {Looser}, {Luetzgendorf}, {Maseda}, {Rawle}, {Rix}, {Rodriguez Del Pino}, {Alberts}, {Egami}, {Eisenstein}, {Endsley}, {Hainline}, {Hausen}, {Johnson}, {Rieke}, {Rieke}, {Robertson}, {Shivaei}, {Stark}, {Sun}, {Tacchella}, {Tang}, {Williams}, {Willmer}, {Baker}, {Baum}, {Bhatawdekar}, {Bowler}, {Boyett}, {Chen}, {Circosta}, {Helton}, {Ji}, {Lyu}, {Nelson}, {Parlanti}, {Perna}, {Sandles}, {Scholtz}, {Suess}, {Topping}, {Uebler}, {Wallace}, \& {Whitler}}]{Bunker23}
{Bunker}, A.~J., {Saxena}, A., {Cameron}, A.~J., {et~al.} 2023, arXiv e-prints, arXiv:2302.07256, \dodoi{10.48550/arXiv.2302.07256}

\bibitem[{{Cain} {et~al.}(2025){Cain}, {D'Aloisio}, \& {Munoz}}]{Cain25}
{Cain}, C., {D'Aloisio}, A., \& {Munoz}, J.~B. 2025, arXiv e-prints, arXiv:2503.08778, \dodoi{10.48550/arXiv.2503.08778}

\bibitem[{{Carniani} {et~al.}(2024){Carniani}, {Hainline}, {D'Eugenio}, {Eisenstein}, {Jakobsen}, {Witstok}, {Johnson}, {Chevallard}, {Maiolino}, {Helton}, {Willott}, {Robertson}, {Alberts}, {Arribas}, {Baker}, {Bhatawdekar}, {Boyett}, {Bunker}, {Cameron}, {Cargile}, {Charlot}, {Curti}, {Curtis-Lake}, {Egami}, {Giardino}, {Isaak}, {Ji}, {Jones}, {Kumari}, {Maseda}, {Parlanti}, {P{\'e}rez-Gonz{\'a}lez}, {Rawle}, {Rieke}, {Rieke}, {Del Pino}, {Saxena}, {Scholtz}, {Smit}, {Sun}, {Tacchella}, {{\"U}bler}, {Venturi}, {Williams}, \& {Willmer}}]{Carniani24a}
{Carniani}, S., {Hainline}, K., {D'Eugenio}, F., {et~al.} 2024, \nat, 633, 318, \dodoi{10.1038/s41586-024-07860-9}

\bibitem[{{Carr} {et~al.}(2025){Carr}, {Cen}, {Scarlata}, {Xu}, {Henry}, {Marques-Chaves}, {Schaerer}, {Amor{\'\i}n}, {Oey}, {Komarova}, {Flury}, {Jaskot}, {Saldana-Lopez}, {Ji}, {Huberty}, {Heckman}, {{\"O}stlin}, {Bait}, {Hayes}, {Thuan}, {Ravindranath}, {Berg}, {Giavalisco}, {Rutkowski}, {Borthakur}, {Chisholm}, {Ferguson}, {Michel-Dansac}, {Verhamme}, \& {Worseck}}]{Carr25}
{Carr}, C.~A., {Cen}, R., {Scarlata}, C., {et~al.} 2025, \apj, 982, 137, \dodoi{10.3847/1538-4357/adb72f}

\bibitem[{{Chemerynska} {et~al.}(2024){Chemerynska}, {Atek}, {Furtak}, {Zitrin}, {Greene}, {Dayal}, {Weibel}, {Fujimoto}, {Kokorev}, {Goulding}, {Williams}, {Nanayakkara}, {Bezanson}, {Brammer}, {Cutler}, {Labbe}, {Leja}, {Pan}, {Price}, {van Dokkum}, {Wang}, {Weaver}, \& {Whitaker}}]{Chemerynska24}
{Chemerynska}, I., {Atek}, H., {Furtak}, L.~J., {et~al.} 2024, \mnras, 531, 2615, \dodoi{10.1093/mnras/stae1260}

\bibitem[{Chen {et~al.}(2024)Chen, Motohara, Spitler, Nakajima, \& Terao}]{Chen24}
Chen, N., Motohara, K., Spitler, L., Nakajima, K., \& Terao, Y. 2024, The Astrophysical Journal, 968, 32, \dodoi{10.3847/1538-4357/ad4033}

\bibitem[{{Chisholm} {et~al.}(2022){Chisholm}, {Saldana-Lopez}, {Flury}, {Schaerer}, {Jaskot}, {Amor{\'\i}n}, {Atek}, {Finkelstein}, {Fleming}, {Ferguson}, {Fern{\'a}ndez}, {Giavalisco}, {Hayes}, {Heckman}, {Henry}, {Ji}, {Marques-Chaves}, {Mauerhofer}, {McCandliss}, {Oey}, {{\"O}stlin}, {Rutkowski}, {Scarlata}, {Thuan}, {Trebitsch}, {Wang}, {Worseck}, \& {Xu}}]{Chisholm22}
{Chisholm}, J., {Saldana-Lopez}, A., {Flury}, S., {et~al.} 2022, \mnras, 517, 5104, \dodoi{10.1093/mnras/stac2874}

\bibitem[{{Cole} {et~al.}(2025){Cole}, {Papovich}, {Finkelstein}, {Bagley}, {Dickinson}, {Iyer}, {Yung}, {Ciesla}, {Amor{\'\i}n}, {Arrabal Haro}, {Bhatawdekar}, {Calabr{\`o}}, {Cleri}, {de la Vega}, {Dekel}, {Endsley}, {Gawiser}, {Giavalisco}, {Hathi}, {Hirschmann}, {Holwerda}, {Kartaltepe}, {Koekemoer}, {Lucas}, {Mascia}, {Mobasher}, {P{\'e}rez-Gonz{\'a}lez}, {Rodighiero}, {Ronayne}, {Tacchella}, {Weiner}, \& {Wilkins}}]{Cole25}
{Cole}, J.~W., {Papovich}, C., {Finkelstein}, S.~L., {et~al.} 2025, \apj, 979, 193, \dodoi{10.3847/1538-4357/ad9a6a}

\bibitem[{{Cullen} {et~al.}(2024){Cullen}, {McLeod}, {McLure}, {Dunlop}, {Donnan}, {Carnall}, {Keating}, {Magee}, {Arellano-Cordova}, {Bowler}, {Begley}, {Flury}, {Hamadouche}, \& {Stanton}}]{Cullen24}
{Cullen}, F., {McLeod}, D.~J., {McLure}, R.~J., {et~al.} 2024, \mnras, 531, 997, \dodoi{10.1093/mnras/stae1211}

\bibitem[{{Curti} {et~al.}(2024){Curti}, {Maiolino}, {Curtis-Lake}, {Chevallard}, {Carniani}, {D'Eugenio}, {Looser}, {Scholtz}, {Charlot}, {Cameron}, {{\"U}bler}, {Witstok}, {Boyett}, {Laseter}, {Sandles}, {Arribas}, {Bunker}, {Giardino}, {Maseda}, {Rawle}, {Rodr{\'\i}guez Del Pino}, {Smit}, {Willott}, {Eisenstein}, {Hausen}, {Johnson}, {Rieke}, {Robertson}, {Tacchella}, {Williams}, {Willmer}, {Baker}, {Bhatawdekar}, {Egami}, {Helton}, {Ji}, {Kumari}, {Perna}, {Shivaei}, \& {Sun}}]{Curti24}
{Curti}, M., {Maiolino}, R., {Curtis-Lake}, E., {et~al.} 2024, \aap, 684, A75, \dodoi{10.1051/0004-6361/202346698}

\bibitem[{{Curtis-Lake} {et~al.}(2023){Curtis-Lake}, {Carniani}, {Cameron}, {Charlot}, {Jakobsen}, {Maiolino}, {Bunker}, {Witstok}, {Smit}, {Chevallard}, {Willott}, {Ferruit}, {Arribas}, {Bonaventura}, {Curti}, {D'Eugenio}, {Franx}, {Giardino}, {Looser}, {L{\"u}tzgendorf}, {Maseda}, {Rawle}, {Rix}, {Rodr{\'\i}guez del Pino}, {{\"U}bler}, {Sirianni}, {Dressler}, {Egami}, {Eisenstein}, {Endsley}, {Hainline}, {Hausen}, {Johnson}, {Rieke}, {Robertson}, {Shivaei}, {Stark}, {Tacchella}, {Williams}, {Willmer}, {Bhatawdekar}, {Bowler}, {Boyett}, {Chen}, {de Graaff}, {Helton}, {Hviding}, {Jones}, {Kumari}, {Lyu}, {Nelson}, {Perna}, {Sandles}, {Saxena}, {Suess}, {Sun}, {Topping}, {Wallace}, \& {Whitler}}]{Curtis23}
{Curtis-Lake}, E., {Carniani}, S., {Cameron}, A., {et~al.} 2023, Nature Astronomy, \dodoi{10.1038/s41550-023-01918-w}

\bibitem[{{Dayal} \& {Ferrara}(2018)}]{Dayal18}
{Dayal}, P., \& {Ferrara}, A. 2018, \physrep, 780, 1, \dodoi{10.1016/j.physrep.2018.10.002}

\bibitem[{{Decataldo} {et~al.}(2017){Decataldo}, {Ferrara}, {Pallottini}, {Gallerani}, \& {Vallini}}]{decataldo:2017}
{Decataldo}, D., {Ferrara}, A., {Pallottini}, A., {Gallerani}, S., \& {Vallini}, L. 2017, \mnras, 471, 4476, \dodoi{10.1093/mnras/stx1879}

\bibitem[{{Donnan} {et~al.}(2023){Donnan}, {McLeod}, {Dunlop}, {McLure}, {Carnall}, {Begley}, {Cullen}, {Hamadouche}, {Bowler}, {Magee}, {McCracken}, {Milvang-Jensen}, {Moneti}, \& {Targett}}]{Donnan23a}
{Donnan}, C.~T., {McLeod}, D.~J., {Dunlop}, J.~S., {et~al.} 2023, \mnras, 518, 6011, \dodoi{10.1093/mnras/stac3472}

\bibitem[{{Dottorini} {et~al.}(2024){Dottorini}, {Calabr{\`o}}, {Pentericci}, {Mascia}, {Llerena}, {Napolitano}, {Santini}, {Roberts-Borsani}, {Castellano}, {Amor{\'\i}n}, {Dickinson}, {Fontana}, {Hathi}, {Hirschmann}, {Koekemoer}, {Lucas}, {Merlin}, {Morales}, {Pacucci}, {Wilkins}, {Arrabal Haro}, {Bagley}, {Finkelstein}, {Kartaltepe}, {Papovich}, \& {Pirzkal}}]{Dottorini24}
{Dottorini}, D., {Calabr{\`o}}, A., {Pentericci}, L., {et~al.} 2024, arXiv e-prints, arXiv:2412.01623, \dodoi{10.48550/arXiv.2412.01623}

\bibitem[{{Enders} {et~al.}(2023){Enders}, {Bomans}, \& {Wittje}}]{Enders23}
{Enders}, A.~U., {Bomans}, D.~J., \& {Wittje}, A. 2023, \aap, 672, A11, \dodoi{10.1051/0004-6361/202245167}

\bibitem[{{Endsley} {et~al.}(2024){Endsley}, {Stark}, {Whitler}, {Topping}, {Johnson}, {Robertson}, {Tacchella}, {Alberts}, {Baker}, {Bhatawdekar}, {Boyett}, {Bunker}, {Cameron}, {Carniani}, {Charlot}, {Chen}, {Chevallard}, {Curtis-Lake}, {Danhaive}, {Egami}, {Eisenstein}, {Hainline}, {Helton}, {Ji}, {Looser}, {Maiolino}, {Nelson}, {Pusk{\'a}s}, {Rieke}, {Rieke}, {Rix}, {Sandles}, {Saxena}, {Simmonds}, {Smit}, {Sun}, {Williams}, {Willmer}, {Willott}, \& {Witstok}}]{Endsley24}
{Endsley}, R., {Stark}, D.~P., {Whitler}, L., {et~al.} 2024, \mnras, 533, 1111, \dodoi{10.1093/mnras/stae1857}

\bibitem[{{Ferrara}(2024{\natexlab{a}})}]{Ferrara24a}
{Ferrara}, A. 2024{\natexlab{a}}, \aap, 684, A207, \dodoi{10.1051/0004-6361/202348321}

\bibitem[{{Ferrara}(2024{\natexlab{b}})}]{Ferrara24b}
---. 2024{\natexlab{b}}, \aap, 689, A310, \dodoi{10.1051/0004-6361/202450944}

\bibitem[{{Ferrara} {et~al.}(2025{\natexlab{a}}){Ferrara}, {Carniani}, {di Mascia}, {Bouwens}, {Oesch}, \& {Schouws}}]{Ferrara25a}
{Ferrara}, A., {Carniani}, S., {di Mascia}, F., {et~al.} 2025{\natexlab{a}}, \aap, 694, A215, \dodoi{10.1051/0004-6361/202452368}

\bibitem[{{Ferrara} {et~al.}(2023){Ferrara}, {Pallottini}, \& {Dayal}}]{Ferrara23}
{Ferrara}, A., {Pallottini}, A., \& {Dayal}, P. 2023, \mnras, 522, 3986, \dodoi{10.1093/mnras/stad1095}

\bibitem[{{Ferrara} {et~al.}(2025{\natexlab{b}}){Ferrara}, {Pallottini}, \& {Sommovigo}}]{Ferrara25b}
{Ferrara}, A., {Pallottini}, A., \& {Sommovigo}, L. 2025{\natexlab{b}}, \aap, 694, A286, \dodoi{10.1051/0004-6361/202452707}

\bibitem[{{Finkelstein} {et~al.}(2024){Finkelstein}, {Leung}, {Bagley}, {Dickinson}, {Ferguson}, {Papovich}, {Akins}, {Arrabal Haro}, {Dav{\'e}}, {Dekel}, {Kartaltepe}, {Kocevski}, {Koekemoer}, {Pirzkal}, {Somerville}, {Yung}, {Amor{\'\i}n}, {Backhaus}, {Behroozi}, {Bisigello}, {Bromm}, {Casey}, {Ch{\'a}vez Ortiz}, {Cheng}, {Chworowsky}, {Cleri}, {Cooper}, {Davis}, {de la Vega}, {Elbaz}, {Franco}, {Fontana}, {Fujimoto}, {Giavalisco}, {Grogin}, {Holwerda}, {Huertas-Company}, {Hirschmann}, {Iyer}, {Jogee}, {Jung}, {Larson}, {Lucas}, {Mobasher}, {Morales}, {Morley}, {Mukherjee}, {P{\'e}rez-Gonz{\'a}lez}, {Ravindranath}, {Rodighiero}, {Rowland}, {Tacchella}, {Taylor}, {Trump}, \& {Wilkins}}]{Finkelstein24}
{Finkelstein}, S.~L., {Leung}, G. C.~K., {Bagley}, M.~B., {et~al.} 2024, \apjl, 969, L2, \dodoi{10.3847/2041-8213/ad4495}

\bibitem[{{Fiore} {et~al.}(2023){Fiore}, {Ferrara}, {Bischetti}, {Feruglio}, \& {Travascio}}]{Fiore23}
{Fiore}, F., {Ferrara}, A., {Bischetti}, M., {Feruglio}, C., \& {Travascio}, A. 2023, \apjl, 943, L27, \dodoi{10.3847/2041-8213/acb5f2}

\bibitem[{{Flury} {et~al.}(2024{\natexlab{a}}){Flury}, {Jaskot}, \& {Silveyra}}]{Flury24}
{Flury}, S., {Jaskot}, A., \& {Silveyra}, A. 2024{\natexlab{a}}, {LyCsurv}, v0.1.0,  Zenodo, \dodoi{10.5281/zenodo.11392442}

\bibitem[{{Flury} {et~al.}(2023){Flury}, {Moran}, \& {Eleazer}}]{Flury23}
{Flury}, S.~R., {Moran}, E.~C., \& {Eleazer}, M. 2023, \mnras, 525, 4231, \dodoi{10.1093/mnras/stad2421}

\bibitem[{{Flury} {et~al.}(2022){Flury}, {Jaskot}, {Ferguson}, {Worseck}, {Makan}, {Chisholm}, {Saldana-Lopez}, {Schaerer}, {McCandliss}, {Xu}, {Wang}, {Oey}, {Ford}, {Heckman}, {Ji}, {Giavalisco}, {Amor{\'\i}n}, {Atek}, {Blaizot}, {Borthakur}, {Carr}, {Castellano}, {De Barros}, {Dickinson}, {Finkelstein}, {Fleming}, {Fontanot}, {Garel}, {Grazian}, {Hayes}, {Henry}, {Mauerhofer}, {Micheva}, {Ostlin}, {Papovich}, {Pentericci}, {Ravindranath}, {Rosdahl}, {Rutkowski}, {Santini}, {Scarlata}, {Teplitz}, {Thuan}, {Trebitsch}, {Vanzella}, \& {Verhamme}}]{Flury22}
{Flury}, S.~R., {Jaskot}, A.~E., {Ferguson}, H.~C., {et~al.} 2022, \apj, 930, 126, \dodoi{10.3847/1538-4357/ac61e4}

\bibitem[{{Flury} {et~al.}(2024{\natexlab{b}}){Flury}, {Jaskot}, {Saldana-Lopez}, {Oey}, {Chisholm}, {Amor{\'\i}n}, {Bait}, {Borthakur}, {Carr}, {Ferguson}, {Giavalisco}, {Hayes}, {Heckman}, {Henry}, {Ji}, {Komarova}, {Leclercq}, {Le Reste}, {McCandliss}, {Marques-Chaves}, {{\"O}stlin}, {Pentericci}, {Ravindranath}, {Rutkowski}, {Scarlata}, {Schaerer}, {Thuan}, {Trebitsch}, {Vanzella}, {Verhamme}, {Wang}, {Worseck}, \& {Xu}}]{Flury24a}
{Flury}, S.~R., {Jaskot}, A.~E., {Saldana-Lopez}, A., {et~al.} 2024{\natexlab{b}}, arXiv e-prints, arXiv:2409.12118, \dodoi{10.48550/arXiv.2409.12118}

\bibitem[{{Foreman-Mackey} {et~al.}(2013){Foreman-Mackey}, {Hogg}, {Lang}, \& {Goodman}}]{EMCEE13}
{Foreman-Mackey}, D., {Hogg}, D.~W., {Lang}, D., \& {Goodman}, J. 2013, \pasp, 125, 306, \dodoi{10.1086/670067}

\bibitem[{{Fujimoto} {et~al.}(2023){Fujimoto}, {Wang}, {Weaver}, {Kokorev}, {Atek}, {Bezanson}, {Labbe}, {Brammer}, {Greene}, {Chemerynska}, {Dayal}, {de Graaff}, {Furtak}, {Oesch}, {Setton}, {Price}, {Miller}, {Williams}, {Whitaker}, {Zitrin}, {Cutler}, {Leja}, {Pan}, {Coe}, {van Dokkum}, {Feldmann}, {Fudamoto}, {Goulding}, {Khullar}, {Marchesini}, {Maseda}, {Nanayakkara}, {Nelson}, {Smit}, {Stefanon}, \& {Weibel}}]{Fujimoto23b}
{Fujimoto}, S., {Wang}, B., {Weaver}, J., {et~al.} 2023, arXiv e-prints, arXiv:2308.11609, \dodoi{10.48550/arXiv.2308.11609}

\bibitem[{{Gelli} {et~al.}(2025){Gelli}, {Pallottini}, {Salvadori}, {Ferrara}, {Mason}, {Carniani}, \& {Ginolfi}}]{Gelli25}
{Gelli}, V., {Pallottini}, A., {Salvadori}, S., {et~al.} 2025, arXiv e-prints, arXiv:2501.16418, \dodoi{10.48550/arXiv.2501.16418}

\bibitem[{{Gelli} {et~al.}(2023){Gelli}, {Salvadori}, {Ferrara}, {Pallottini}, \& {Carniani}}]{Gelli23}
{Gelli}, V., {Salvadori}, S., {Ferrara}, A., {Pallottini}, A., \& {Carniani}, S. 2023, \apjl, 954, L11, \dodoi{10.3847/2041-8213/acee80}

\bibitem[{{Gnedin}(2000)}]{Gnedin00}
{Gnedin}, N.~Y. 2000, \apj, 542, 535, \dodoi{10.1086/317042}

\bibitem[{{Gnedin} \& {Madau}(2022)}]{Gnedin22}
{Gnedin}, N.~Y., \& {Madau}, P. 2022, Living Reviews in Computational Astrophysics, 8, 3, \dodoi{10.1007/s41115-022-00015-5}

\bibitem[{{Goodman} \& {Weare}(2010)}]{Goodman10}
{Goodman}, J., \& {Weare}, J. 2010, Communications in Applied Mathematics and Computational Science, 5, 65, \dodoi{10.2140/camcos.2010.5.65}

\bibitem[{{Grazian} {et~al.}(2017){Grazian}, {Giallongo}, {Paris}, {Boutsia}, {Dickinson}, {Santini}, {Windhorst}, {Jansen}, {Cohen}, {Ashcraft}, {Scarlata}, {Rutkowski}, {Vanzella}, {Cusano}, {Cristiani}, {Giavalisco}, {Ferguson}, {Koekemoer}, {Grogin}, {Castellano}, {Fiore}, {Fontana}, {Marchi}, {Pedichini}, {Pentericci}, {Amor{\'\i}n}, {Barro}, {Bonchi}, {Bongiorno}, {Faber}, {Fumana}, {Galametz}, {Guaita}, {Kocevski}, {Merlin}, {Nonino}, {O'Connell}, {Pilo}, {Ryan}, {Sani}, {Speziali}, {Testa}, {Weiner}, \& {Yan}}]{Grazian17}
{Grazian}, A., {Giallongo}, E., {Paris}, D., {et~al.} 2017, \aap, 602, A18, \dodoi{10.1051/0004-6361/201730447}

\bibitem[{{Hainline} {et~al.}(2024){Hainline}, {D'Eugenio}, {Jakobsen}, {Chevallard}, {Carniani}, {Witstok}, {Ji}, {Curtis-Lake}, {Johnson}, {Robertson}, {Tacchella}, {Curti}, {Charlot}, {Helton}, {Arribas}, {Bhatawdekar}, {Bunker}, {Cameron}, {Egami}, {Eisenstein}, {Hausen}, {Kumari}, {Maiolino}, {P{\'e}rez-Gonz{\'a}lez}, {Rieke}, {Saxena}, {Scholtz}, {Smit}, {Sun}, {Williams}, {Willmer}, \& {Willott}}]{Hainline24}
{Hainline}, K.~N., {D'Eugenio}, F., {Jakobsen}, P., {et~al.} 2024, \apj, 976, 160, \dodoi{10.3847/1538-4357/ad8447}

\bibitem[{{Harikane} {et~al.}(2024){Harikane}, {Nakajima}, {Ouchi}, {Umeda}, {Isobe}, {Ono}, {Xu}, \& {Zhang}}]{Harikane23}
{Harikane}, Y., {Nakajima}, K., {Ouchi}, M., {et~al.} 2024, \apj, 960, 56, \dodoi{10.3847/1538-4357/ad0b7e}

\bibitem[{{Helton} {et~al.}(2024){Helton}, {Rieke}, {Alberts}, {Wu}, {Eisenstein}, {Hainline}, {Carniani}, {Ji}, {Baker}, {Bhatawdekar}, {Bunker}, {Cargile}, {Charlot}, {Chevallard}, {D'Eugenio}, {Egami}, {Johnson}, {Jones}, {Lyu}, {Maiolino}, {P{\'e}rez-Gonz{\'a}lez}, {Rieke}, {Robertson}, {Saxena}, {Scholtz}, {Shivaei}, {Sun}, {Tacchella}, {Whitler}, {Williams}, {Willmer}, {Willott}, {Witstok}, \& {Zhu}}]{Helton24}
{Helton}, J.~M., {Rieke}, G.~H., {Alberts}, S., {et~al.} 2024, arXiv e-prints, arXiv:2405.18462, \dodoi{10.48550/arXiv.2405.18462}

\bibitem[{{Hogarth} {et~al.}(2020){Hogarth}, {Amor{\'\i}n}, {V{\'\i}lchez}, {H{\"a}gele}, {Cardaci}, {P{\'e}rez-Montero}, {Firpo}, {Jaskot}, \& {Ch{\'a}vez}}]{Hogarth20}
{Hogarth}, L., {Amor{\'\i}n}, R., {V{\'\i}lchez}, J.~M., {et~al.} 2020, \mnras, 494, 3541, \dodoi{10.1093/mnras/staa851}

\bibitem[{{Hunter}(2007)}]{Hunter07}
{Hunter}, J.~D. 2007, Computing in Science and Engineering, 9, 90, \dodoi{10.1109/MCSE.2007.55}

\bibitem[{{Inoue} {et~al.}(2014){Inoue}, {Shimizu}, {Iwata}, \& {Tanaka}}]{Inoue14}
{Inoue}, A.~K., {Shimizu}, I., {Iwata}, I., \& {Tanaka}, M. 2014, \mnras, 442, 1805, \dodoi{10.1093/mnras/stu936}

\bibitem[{{Izotov} {et~al.}(2018{\natexlab{a}}){Izotov}, {Schaerer}, {Worseck}, {Guseva}, {Thuan}, {Verhamme}, {Orlitov{\'a}}, \& {Fricke}}]{Izotov18b}
{Izotov}, Y.~I., {Schaerer}, D., {Worseck}, G., {et~al.} 2018{\natexlab{a}}, \mnras, 474, 4514, \dodoi{10.1093/mnras/stx3115}

\bibitem[{{Izotov} {et~al.}(2021){Izotov}, {Worseck}, {Schaerer}, {Guseva}, {Chisholm}, {Thuan}, {Fricke}, \& {Verhamme}}]{Izotov21}
{Izotov}, Y.~I., {Worseck}, G., {Schaerer}, D., {et~al.} 2021, \mnras, 503, 1734, \dodoi{10.1093/mnras/stab612}

\bibitem[{{Izotov} {et~al.}(2018{\natexlab{b}}){Izotov}, {Worseck}, {Schaerer}, {Guseva}, {Thuan}, {Fricke}, \& {Orlitov{\'a}}}]{Izotov18}
---. 2018{\natexlab{b}}, \mnras, 478, 4851, \dodoi{10.1093/mnras/sty1378}

\bibitem[{{Jaskot} {et~al.}(2024){Jaskot}, {Silveyra}, {Plantinga}, {Flury}, {Hayes}, {Chisholm}, {Heckman}, {Pentericci}, {Schaerer}, {Trebitsch}, {Verhamme}, {Carr}, {Ferguson}, {Ji}, {Giavalisco}, {Henry}, {Marques-Chaves}, {{\"O}stlin}, {Saldana-Lopez}, {Scarlata}, {Worseck}, \& {Xu}}]{Jaskot24}
{Jaskot}, A.~E., {Silveyra}, A.~C., {Plantinga}, A., {et~al.} 2024, \apj, 973, 111, \dodoi{10.3847/1538-4357/ad5557}

\bibitem[{{Jecmen} \& {Oey}(2023)}]{Jecmen23}
{Jecmen}, M.~C., \& {Oey}, M.~S. 2023, \apj, 958, 149, \dodoi{10.3847/1538-4357/ad0460}

\bibitem[{{Ji} {et~al.}(2025){Ji}, {Alberts}, {Zhu}, {Vanzella}, {Giavalisco}, {Hainline}, {Baker}, {Bunker}, {Helton}, {Lyu}, {Rinaldi}, {Robertson}, {Simmonds}, {Tacchella}, {Williams}, {Willmer}, \& {Witstok}}]{Ji25}
{Ji}, Z., {Alberts}, S., {Zhu}, Y., {et~al.} 2025, arXiv e-prints, arXiv:2504.01067, \dodoi{10.48550/arXiv.2504.01067}

\bibitem[{{Kimm} {et~al.}(2019){Kimm}, {Blaizot}, {Garel}, {Michel-Dansac}, {Katz}, {Rosdahl}, {Verhamme}, \& {Haehnelt}}]{Kimm19}
{Kimm}, T., {Blaizot}, J., {Garel}, T., {et~al.} 2019, \mnras, 486, 2215, \dodoi{10.1093/mnras/stz989}

\bibitem[{{Kimm} {et~al.}(2018){Kimm}, {Haehnelt}, {Blaizot}, {Katz}, {Michel-Dansac}, {Garel}, {Rosdahl}, \& {Teyssier}}]{Kimm18}
{Kimm}, T., {Haehnelt}, M., {Blaizot}, J., {et~al.} 2018, \mnras, 475, 4617, \dodoi{10.1093/mnras/sty126}

\bibitem[{{Kobayashi} \& {Ferrara}(2023)}]{Kobayashi23}
{Kobayashi}, C., \& {Ferrara}, A. 2023, arXiv e-prints, arXiv:2308.15583, \dodoi{10.48550/arXiv.2308.15583}

\bibitem[{{Kokorev} {et~al.}(2025){Kokorev}, {Ch{\'a}vez Ortiz}, {Taylor}, {Finkelstein}, {Arrabal Haro}, {Dickinson}, {Chisholm}, {Fujimoto}, {Mu{\~n}oz}, {Endsley}, {Hu}, {Napolitano}, {Wilkins}, {Akins}, {Amori{\'\i}n}, {Casey}, {Cheng}, {Cleri}, {Cole}, {Cullen}, {Daddi}, {Davis}, {Donnan}, {Dunlop}, {Fern{\'a}ndez}, {Giavalisco}, {Grogin}, {Hathi}, {Hirschmann}, {Kartaltepe}, {Koekemoer}, {Leung}, {Lucas}, {McLeod}, {Papovich}, {Pentericci}, {P{\'e}rez-Gonz{\'a}lez}, {Somerville}, {Wang}, {Yung}, \& {Zavala}}]{Kokorev25}
{Kokorev}, V., {Ch{\'a}vez Ortiz}, {\'O}.~A., {Taylor}, A.~J., {et~al.} 2025, arXiv e-prints, arXiv:2504.12504, \dodoi{10.48550/arXiv.2504.12504}

\bibitem[{{Komarova} {et~al.}(2021){Komarova}, {Oey}, {Krumholz}, {Silich}, {Kumari}, \& {James}}]{Komarova21}
{Komarova}, L., {Oey}, M.~S., {Krumholz}, M.~R., {et~al.} 2021, \apjl, 920, L46, \dodoi{10.3847/2041-8213/ac2c09}

\bibitem[{{Kostyuk} {et~al.}(2025){Kostyuk}, {Ciardi}, \& {Ferrara}}]{Kostyuk25}
{Kostyuk}, I., {Ciardi}, B., \& {Ferrara}, A. 2025, \aap, 695, A32, \dodoi{10.1051/0004-6361/202449997}

\bibitem[{{Kurczynski} {et~al.}(2014){Kurczynski}, {Gawiser}, {Rafelski}, {Teplitz}, {Acquaviva}, {Brown}, {Coe}, {de Mello}, {Finkelstein}, {Grogin}, {Koekemoer}, {Lee}, {Scarlata}, \& {Siana}}]{Kurczynski14}
{Kurczynski}, P., {Gawiser}, E., {Rafelski}, M., {et~al.} 2014, \apjl, 793, L5, \dodoi{10.1088/2041-8205/793/1/L5}

\bibitem[{{Leclercq} {et~al.}(2024){Leclercq}, {Chisholm}, {King}, {Zeimann}, {Jaskot}, {Henry}, {Hayes}, {Flury}, {Izotov}, {Prochaska}, {Verhamme}, {Amor{\'\i}n}, {Atek}, {Bait}, {Blaizot}, {Carr}, {Ji}, {Le Reste}, {Ferguson}, {Gazagnes}, {Heckman}, {Komarova}, {Marques-Chaves}, {{\"O}stlin}, {Saldana-Lopez}, {Scarlata}, {Schaerer}, {Thuan}, {Trebitsch}, {Worseck}, {Wang}, \& {Xu}}]{Leclercq24}
{Leclercq}, F., {Chisholm}, J., {King}, W., {et~al.} 2024, \aap, 687, A73, \dodoi{10.1051/0004-6361/202449362}

\bibitem[{{Llerena} {et~al.}(2024){Llerena}, {Pentericci}, {Napolitano}, {Mascia}, {Amor{\'\i}n}, {Calabr{\`o}}, {Castellano}, {Cleri}, {Giavalisco}, {Grogin}, {Hathi}, {Hirschmann}, {Koekemoer}, {Nanayakkara}, {Pacucci}, {Shen}, {Wilkins}, {Yoon}, {Yung}, {Bhatawdekar}, {Lucas}, {Wang}, {Arrabal Haro}, {Bagley}, {Finkelstein}, {Kartaltepe}, {Merlin}, {Papovich}, \& {Pirzkal}}]{LLerena24}
{Llerena}, M., {Pentericci}, L., {Napolitano}, L., {et~al.} 2024, arXiv e-prints, arXiv:2412.01358, \dodoi{10.48550/arXiv.2412.01358}

\bibitem[{{Looser} {et~al.}(2023){Looser}, {D'Eugenio}, {Maiolino}, {Witstok}, {Sandles}, {Curtis-Lake}, {Chevallard}, {Tacchella}, {Johnson}, {Baker}, {Suess}, {Carniani}, {Ferruit}, {Arribas}, {Bonaventura}, {Bunker}, {Cameron}, {Charlot}, {Curti}, {de Graaff}, {Maseda}, {Rawle}, {Rix}, {Rodriguez Del Pino}, {Smit}, {{\"U}bler}, {Willott}, {Alberts}, {Egami}, {Eisenstein}, {Endsley}, {Hausen}, {Rieke}, {Robertson}, {Shivaei}, {Williams}, {Boyett}, {Chen}, {Ji}, {Jones}, {Kumari}, {Nelson}, {Perna}, {Saxena}, \& {Scholtz}}]{Looser23}
{Looser}, T.~J., {D'Eugenio}, F., {Maiolino}, R., {et~al.} 2023, arXiv e-prints, arXiv:2302.14155.
\newblock \doarXiv{2302.14155}

\bibitem[{{Mainali} {et~al.}(2022){Mainali}, {Rigby}, {Chisholm}, {Bayliss}, {Bordoloi}, {Gladders}, {Rivera-Thorsen}, {Dahle}, {Sharon}, {Florian}, {Berg}, {Sharma}, {Owens}, {Kjellgren}, {Kim}, \& {Wayne}}]{Mainali22}
{Mainali}, R., {Rigby}, J.~R., {Chisholm}, J., {et~al.} 2022, \apj, 940, 160, \dodoi{10.3847/1538-4357/ac9cd6}

\bibitem[{{Marchi} {et~al.}(2017){Marchi}, {Pentericci}, {Guaita}, {Ribeiro}, {Castellano}, {Schaerer}, {Hathi}, {Lemaux}, {Grazian}, {Le F{\`e}vre}, {Garilli}, {Maccagni}, {Amorin}, {Bardelli}, {Cassata}, {Fontana}, {Koekemoer}, {Le Brun}, {Tasca}, {Thomas}, {Vanzella}, {Zamorani}, \& {Zucca}}]{Marchi17}
{Marchi}, F., {Pentericci}, L., {Guaita}, L., {et~al.} 2017, \aap, 601, A73, \dodoi{10.1051/0004-6361/201630054}

\bibitem[{{Marques-Chaves} {et~al.}(2022){Marques-Chaves}, {Schaerer}, {{\'A}lvarez-M{\'a}rquez}, {Verhamme}, {Ceverino}, {Chisholm}, {Colina}, {Dessauges-Zavadsky}, {P{\'e}rez-Fournon}, {Saldana-Lopez}, {Upadhyaya}, \& {Vanzella}}]{Marques22}
{Marques-Chaves}, R., {Schaerer}, D., {{\'A}lvarez-M{\'a}rquez}, J., {et~al.} 2022, \mnras, 517, 2972, \dodoi{10.1093/mnras/stac2893}

\bibitem[{{Marques-Chaves} {et~al.}(2024){Marques-Chaves}, {Schaerer}, {Vanzella}, {Verhamme}, {Dessauges-Zavadsky}, {Chisholm}, {Leclercq}, {Upadhyaya}, {{\'A}lvarez-M{\'a}rquez}, {Colina}, {Garel}, \& {Messa}}]{Marques24}
{Marques-Chaves}, R., {Schaerer}, D., {Vanzella}, E., {et~al.} 2024, \aap, 691, A87, \dodoi{10.1051/0004-6361/202451667}

\bibitem[{{Mascia} {et~al.}(2024){Mascia}, {Pentericci}, {Calabr{\`o}}, {Santini}, {Napolitano}, {Arrabal Haro}, {Castellano}, {Dickinson}, {Ocvirk}, {Lewis}, {Amor{\'\i}n}, {Bagley}, {Bhatawdekar}, {Cleri}, {Costantin}, {Dekel}, {Finkelstein}, {Fontana}, {Giavalisco}, {Grogin}, {Hathi}, {Hirschmann}, {Holwerda}, {Jung}, {Kartaltepe}, {Koekemoer}, {Lucas}, {Papovich}, {P{\'e}rez-Gonz{\'a}lez}, {Pirzkal}, {Trump}, {Wilkins}, \& {Yung}}]{Mascia24}
{Mascia}, S., {Pentericci}, L., {Calabr{\`o}}, A., {et~al.} 2024, \aap, 685, A3, \dodoi{10.1051/0004-6361/202347884}

\bibitem[{{Mascia} {et~al.}(2025){Mascia}, {Pentericci}, {Llerena}, {Calabr{\`o}}, {Matthee}, {Flury}, {Pacucci}, {Jaskot}, {Amor{\'\i}n}, {Bhatawdekar}, {Castellano}, {Cleri}, {Costantin}, {Davis}, {Di Cesare}, {Dickinson}, {Fontana}, {Guo}, {Giavalisco}, {Holwerda}, {Hu}, {Huertas-Company}, {Jung}, {Kartaltepe}, {Kashino}, {Koekemoer}, {Lucas}, {Lotz}, {Napolitano}, {Jogee}, \& {Wilkins}}]{Mascia25}
{Mascia}, S., {Pentericci}, L., {Llerena}, M., {et~al.} 2025, arXiv e-prints, arXiv:2501.08268, \dodoi{10.48550/arXiv.2501.08268}

\bibitem[{{McLeod} {et~al.}(2023){McLeod}, {Donnan}, {McLure}, {Dunlop}, {Magee}, {Begley}, {Carnall}, {Cullen}, {Ellis}, {Hamadouche}, \& {Stanton}}]{McLeod23}
{McLeod}, D.~J., {Donnan}, C.~T., {McLure}, R.~J., {et~al.} 2023, arXiv e-prints, arXiv:2304.14469, \dodoi{10.48550/arXiv.2304.14469}

\bibitem[{{Meiksin}(2009)}]{Meiksin09}
{Meiksin}, A.~A. 2009, Reviews of Modern Physics, 81, 1405, \dodoi{10.1103/RevModPhys.81.1405}

\bibitem[{{Me{\v{s}}tri{\'c}} {et~al.}(2025){Me{\v{s}}tri{\'c}}, {Vanzella}, {Beckett}, {Rafelski}, {Grillo}, {Giavalisco}, {Messa}, {Castellano}, {Calura}, {Cupani}, {Zanella}, {Bergamini}, {Meneghetti}, {Mercurio}, {Rosati}, {Nonino}, {Caputi}, \& {Comastri}}]{Mestric25}
{Me{\v{s}}tri{\'c}}, U., {Vanzella}, E., {Beckett}, A., {et~al.} 2025, arXiv e-prints, arXiv:2504.18711, \dodoi{10.48550/arXiv.2504.18711}

\bibitem[{{Mirocha} \& {Furlanetto}(2023)}]{Mirocha23}
{Mirocha}, J., \& {Furlanetto}, S.~R. 2023, \mnras, 519, 843, \dodoi{10.1093/mnras/stac3578}

\bibitem[{{Morales} {et~al.}(2024{\natexlab{a}}){Morales}, {Finkelstein}, {Bagley}, {Alavi}, {Grogin}, {Hathi}, {Koekemoer}, {Nedkova}, {Prichard}, {Rafelski}, {Sunnquist}, {Taamoli}, {Teplitz}, {Wang}, {Windhorst}, \& {Yung}}]{Morales24}
{Morales}, A., {Finkelstein}, S., {Bagley}, M., {et~al.} 2024{\natexlab{a}}, arXiv e-prints, arXiv:2405.20901, \dodoi{10.48550/arXiv.2405.20901}

\bibitem[{{Morales} {et~al.}(2024{\natexlab{b}}){Morales}, {Finkelstein}, {Leung}, {Bagley}, {Cleri}, {Dave}, {Dickinson}, {Ferguson}, {Hathi}, {Jones}, {Koekemoer}, {Papovich}, {P{\'e}rez-Gonz{\'a}lez}, {Pirzkal}, {Smith}, {Wilkins}, \& {Yung}}]{Morales24b}
{Morales}, A.~M., {Finkelstein}, S.~L., {Leung}, G. C.~K., {et~al.} 2024{\natexlab{b}}, \apjl, 964, L24, \dodoi{10.3847/2041-8213/ad2de4}

\bibitem[{{Mu{\~n}oz} {et~al.}(2024){Mu{\~n}oz}, {Mirocha}, {Chisholm}, {Furlanetto}, \& {Mason}}]{Munoz24}
{Mu{\~n}oz}, J.~B., {Mirocha}, J., {Chisholm}, J., {Furlanetto}, S.~R., \& {Mason}, C. 2024, \mnras, 535, L37, \dodoi{10.1093/mnrasl/slae086}

\bibitem[{{Nakajima} {et~al.}(2020){Nakajima}, {Ellis}, {Robertson}, {Tang}, \& {Stark}}]{Nakajima20}
{Nakajima}, K., {Ellis}, R.~S., {Robertson}, B.~E., {Tang}, M., \& {Stark}, D.~P. 2020, \apj, 889, 161, \dodoi{10.3847/1538-4357/ab6604}

\bibitem[{{Nakazato} \& {Ferrara}(2024)}]{Nakazato25}
{Nakazato}, Y., \& {Ferrara}, A. 2024, arXiv e-prints, arXiv:2412.07598, \dodoi{10.48550/arXiv.2412.07598}

\bibitem[{{Nanayakkara} {et~al.}(2023){Nanayakkara}, {Glazebrook}, {Jacobs}, {Bonchi}, {Castellano}, {Fontana}, {Mason}, {Merlin}, {Morishita}, {Paris}, {Trenti}, {Treu}, {Calabr{\`o}}, {Boyett}, {Bradac}, {Leethochawalit}, {Marchesini}, {Santini}, {Strait}, {Vanzella}, {Vulcani}, {Wang}, \& {Yang}}]{Nanayakkara23}
{Nanayakkara}, T., {Glazebrook}, K., {Jacobs}, C., {et~al.} 2023, \apjl, 947, L26, \dodoi{10.3847/2041-8213/acbfb9}

\bibitem[{{Narayanan} {et~al.}(2024){Narayanan}, {Stark}, {Finkelstein}, {Torrey}, {Li}, {Cullen}, {Topping}, {Marinacci}, {Sales}, {Shen}, \& {Vogelsberger}}]{Narayanan24}
{Narayanan}, D., {Stark}, D.~P., {Finkelstein}, S.~L., {et~al.} 2024, arXiv e-prints, arXiv:2408.13312, \dodoi{10.48550/arXiv.2408.13312}

\bibitem[{{Nebrin} {et~al.}(2025){Nebrin}, {Smith}, {Lorinc}, {H{\"o}rnquist}, {Larson}, {Mellema}, \& {Giri}}]{Nebrin25}
{Nebrin}, O., {Smith}, A., {Lorinc}, K., {et~al.} 2025, \mnras, 537, 1646, \dodoi{10.1093/mnras/staf038}

\bibitem[{{Overzier} {et~al.}(2011){Overzier}, {Heckman}, {Wang}, {Armus}, {Buat}, {Howell}, {Meurer}, {Seibert}, {Siana}, {Basu-Zych}, {Charlot}, {Gon{\c{c}}alves}, {Martin}, {Neill}, {Rich}, {Salim}, \& {Schiminovich}}]{Overzier11}
{Overzier}, R.~A., {Heckman}, T.~M., {Wang}, J., {et~al.} 2011, \apjl, 726, L7, \dodoi{10.1088/2041-8205/726/1/L7}

\bibitem[{{Pahl} {et~al.}(2025){Pahl}, {Topping}, {Shapley}, {Sanders}, {Reddy}, {Clarke}, {Kehoe}, {Bento}, \& {Brammer}}]{Pahl25}
{Pahl}, A., {Topping}, M.~W., {Shapley}, A., {et~al.} 2025, \apj, 981, 134, \dodoi{10.3847/1538-4357/adb1ab}

\bibitem[{{Pallottini} \& {Ferrara}(2023)}]{Pallottini23}
{Pallottini}, A., \& {Ferrara}, A. 2023, \aap, 677, L4, \dodoi{10.1051/0004-6361/202347384}

\bibitem[{Papovich {et~al.}(2025)Papovich, Cole, Hu, Finkelstein, Shen, Haro, Amorín, Backhaus, Bagley, Bhatawdekar, Calabró, Carnall, Cleri, Daddi, Dickinson, Grogin, Holwerda, Jaskot, Koekemoer, Llerena, Lucas, Mascia, Pacucci, Pentericci, Pérez-González, Pirzkal, Raghunathan, Seillé, Somerville, \& Yung}]{Papovich25}
Papovich, C., Cole, J.~W., Hu, W., {et~al.} 2025, Galaxies in the Epoch of Reionization Are All Bark and No Bite -- Plenty of Ionizing Photons, Low Escape Fractions.
\newblock \doarXiv{2505.08870}

\bibitem[{{P{\'e}rez-Gonz{\'a}lez} {et~al.}(2025){P{\'e}rez-Gonz{\'a}lez}, {{\"O}stlin}, {Costantin}, {Melinder}, {Finkelstein}, {Somerville}, {Annunziatella}, {{\'A}lvarez-M{\'a}rquez}, {Colina}, {Dekel}, {Ferguson}, {Li}, {Yung}, {Bagley}, {Boogard}, {Burgarella}, {Calabr{\`o}}, {Caputi}, {Cheng}, {Eckart}, {Giavalisco}, {Gillman}, {Greve}, {Hathi}, {Hjorth}, {Huertas-Company}, {Kartaltepe}, {Koekemoer}, {Kokorev}, {Labiano}, {Langeroodi}, {Leung}, {Natarajan}, {Papovich}, {Peissker}, {Pentericci}, {Pirzkal}, {Rinaldi}, {van der Werf}, \& {Walter}}]{Perez25}
{P{\'e}rez-Gonz{\'a}lez}, P.~G., {{\"O}stlin}, G., {Costantin}, L., {et~al.} 2025, arXiv e-prints, arXiv:2503.15594, \dodoi{10.48550/arXiv.2503.15594}

\bibitem[{{Pizzati} {et~al.}(2020){Pizzati}, {Ferrara}, {Pallottini}, {Gallerani}, {Vallini}, {Decataldo}, \& {Fujimoto}}]{Pizzati20}
{Pizzati}, E., {Ferrara}, A., {Pallottini}, A., {et~al.} 2020, \mnras, 495, 160, \dodoi{10.1093/mnras/staa1163}

\bibitem[{{Rivera-Thorsen} {et~al.}(2019){Rivera-Thorsen}, {Dahle}, {Chisholm}, {Florian}, {Gronke}, {Rigby}, {Gladders}, {Mahler}, {Sharon}, \& {Bayliss}}]{Rivera19}
{Rivera-Thorsen}, T.~E., {Dahle}, H., {Chisholm}, J., {et~al.} 2019, Science, 366, 738, \dodoi{10.1126/science.aaw0978}

\bibitem[{{Roberts-Borsani} {et~al.}(2024){Roberts-Borsani}, {Treu}, {Shapley}, {Fontana}, {Pentericci}, {Castellano}, {Morishita}, {Bergamini}, \& {Rosati}}]{Roberts24}
{Roberts-Borsani}, G., {Treu}, T., {Shapley}, A., {et~al.} 2024, \apj, 976, 193, \dodoi{10.3847/1538-4357/ad85d3}

\bibitem[{{Saxena} {et~al.}(2024){Saxena}, {Cameron}, {Katz}, {Bunker}, {Chevallard}, {D'Eugenio}, {Arribas}, {Bhatawdekar}, {Boyett}, {Cargile}, {Carniani}, {Charlot}, {Curti}, {Curtis-Lake}, {Hainline}, {Ji}, {Johnson}, {Jones}, {Kumari}, {Laseter}, {Maseda}, {Robertson}, {Simmonds}, {Tacchella}, {Ubler}, {Williams}, {Willott}, {Witstok}, \& {Zhu}}]{Saxena24}
{Saxena}, A., {Cameron}, A.~J., {Katz}, H., {et~al.} 2024, arXiv e-prints, arXiv:2411.14532, \dodoi{10.48550/arXiv.2411.14532}

\bibitem[{{Shen} {et~al.}(2023){Shen}, {Vogelsberger}, {Boylan-Kolchin}, {Tacchella}, \& {Kannan}}]{Shen23}
{Shen}, X., {Vogelsberger}, M., {Boylan-Kolchin}, M., {Tacchella}, S., \& {Kannan}, R. 2023, arXiv e-prints, arXiv:2305.05679, \dodoi{10.48550/arXiv.2305.05679}

\bibitem[{{Simmonds} {et~al.}(2024){Simmonds}, {Tacchella}, {Hainline}, {Johnson}, {Pusk{\'a}s}, {Robertson}, {Baker}, {Bhatawdekar}, {Boyett}, {Bunker}, {Cargile}, {Carniani}, {Chevallard}, {Curti}, {Curtis-Lake}, {Ji}, {Jones}, {Kumari}, {Laseter}, {Maiolino}, {Maseda}, {Rinaldi}, {Stoffers}, {{\"U}bler}, {Villanueva}, {Williams}, {Willott}, {Witstok}, \& {Zhu}}]{Simmonds24}
{Simmonds}, C., {Tacchella}, S., {Hainline}, K., {et~al.} 2024, \mnras, 535, 2998, \dodoi{10.1093/mnras/stae2537}

\bibitem[{{Smith} {et~al.}(2017){Smith}, {Bromm}, \& {Loeb}}]{Smith17}
{Smith}, A., {Bromm}, V., \& {Loeb}, A. 2017, \mnras, 464, 2963, \dodoi{10.1093/mnras/stw2591}

\bibitem[{{Sukhbold} {et~al.}(2016){Sukhbold}, {Ertl}, {Woosley}, {Brown}, \& {Janka}}]{Sukhbolt16}
{Sukhbold}, T., {Ertl}, T., {Woosley}, S.~E., {Brown}, J.~M., \& {Janka}, H.~T. 2016, \apj, 821, 38, \dodoi{10.3847/0004-637X/821/1/38}

\bibitem[{{Terlevich} {et~al.}(1992){Terlevich}, {Tenorio-Tagle}, {Franco}, \& {Melnick}}]{Terlevich1992}
{Terlevich}, R., {Tenorio-Tagle}, G., {Franco}, J., \& {Melnick}, J. 1992, \mnras, 255, 713, \dodoi{10.1093/mnras/255.4.713}

\bibitem[{{Tomaselli} \& {Ferrara}(2021)}]{Tomaselli21}
{Tomaselli}, G.~M., \& {Ferrara}, A. 2021, \mnras, 504, 89, \dodoi{10.1093/mnras/stab876}

\bibitem[{{Topping} {et~al.}(2022{\natexlab{a}}){Topping}, {Stark}, {Endsley}, {Plat}, {Whitler}, {Chen}, \& {Charlot}}]{Topping_2022}
{Topping}, M.~W., {Stark}, D.~P., {Endsley}, R., {et~al.} 2022{\natexlab{a}}, \apj, 941, 153, \dodoi{10.3847/1538-4357/aca522}

\bibitem[{{Topping} {et~al.}(2022{\natexlab{b}}){Topping}, {Stark}, {Endsley}, {Bouwens}, {Schouws}, {Smit}, {Stefanon}, {Inami}, {Bowler}, {Oesch}, {Gonzalez}, {Dayal}, {da Cunha}, {Algera}, {van der Werf}, {Pallottini}, {Barrufet}, {Schneider}, {De Looze}, {Sommovigo}, {Whitler}, {Graziani}, {Fudamoto}, \& {Ferrara}}]{topping2022}
---. 2022{\natexlab{b}}, \mnras, 516, 975, \dodoi{10.1093/mnras/stac2291}

\bibitem[{{Topping} {et~al.}(2023){Topping}, {Stark}, {Endsley}, {Whitler}, {Hainline}, {Johnson}, {Robertson}, {Tacchella}, {Chen}, {Alberts}, {Baker}, {Bunker}, {Carniani}, {Charlot}, {Chevallard}, {Curtis-Lake}, {DeCoursey}, {Egami}, {Eisenstein}, {Ji}, {Maiolino}, {Williams}, {Willmer}, {Willott}, \& {Witstok}}]{Topping23}
---. 2023, arXiv e-prints, arXiv:2307.08835, \dodoi{10.48550/arXiv.2307.08835}

\bibitem[{{Vanzella} {et~al.}(2016){Vanzella}, {de Barros}, {Vasei}, {Alavi}, {Giavalisco}, {Siana}, {Grazian}, {Hasinger}, {Suh}, {Cappelluti}, {Vito}, {Amorin}, {Balestra}, {Brusa}, {Calura}, {Castellano}, {Comastri}, {Fontana}, {Gilli}, {Mignoli}, {Pentericci}, {Vignali}, \& {Zamorani}}]{Vanzella16}
{Vanzella}, E., {de Barros}, S., {Vasei}, K., {et~al.} 2016, \apj, 825, 41, \dodoi{10.3847/0004-637X/825/1/41}

\bibitem[{{Vanzella} {et~al.}(2022){Vanzella}, {Castellano}, {Bergamini}, {Meneghetti}, {Zanella}, {Calura}, {Caminha}, {Rosati}, {Cupani}, {Me{\v{s}}tri{\'c}}, {Brammer}, {Tozzi}, {Mercurio}, {Grillo}, {Sani}, {Cristiani}, {Nonino}, {Merlin}, \& {Pignataro}}]{Vanzella22}
{Vanzella}, E., {Castellano}, M., {Bergamini}, P., {et~al.} 2022, \aap, 659, A2, \dodoi{10.1051/0004-6361/202141590}

\bibitem[{{Wang} {et~al.}(2025){Wang}, {Teplitz}, {Smith}, {Windhorst}, {Rafelski}, {Mehta}, {Alavi}, {Ji}, {Brammer}, {Colbert}, {Grogin}, {Hathi}, {Koekemoer}, {Prichard}, {Scarlata}, {Sunnquist}, {Arrabal Haro}, {Conselice}, {Gawiser}, {Guo}, {Hayes}, {Jansen}, {Lucas}, {O'Connell}, {Robertson}, {Rutkowski}, {Siana}, {Vanzella}, {Ashcraft}, {Bagley}, {Baronchelli}, {Barro}, {Blanche}, {Broussard}, {Carleton}, {Chartab}, {Cheng}, {Codoreanu}, {Cohen}, {Dai}, {Darvish}, {Dav{\'e}}, {Degroot}, {de Mello}, {Dickinson}, {Emami}, {Ferguson}, {Ferreira}, {Finkelstein}, {Finkelstein}, {Gardner}, {Gburek}, {Giavalisco}, {Grazian}, {Gronwall}, {Hemmati}, {Howell}, {Iyer}, {Kaviraj}, {Kurczynski}, {Lazar}, {MacKenty}, {Mantha}, {Martin}, {Martin}, {McCabe}, {Mobasher}, {Nedkova}, {Olsen}, {Otteson}, {Ravindranath}, {Redshaw}, {Sattari}, {Soto}, {Yung}, {Zabelle}, \& {UVCANDELS Team}}]{Wang25}
{Wang}, X., {Teplitz}, H.~I., {Smith}, B.~M., {et~al.} 2025, \apj, 980, 74, \dodoi{10.3847/1538-4357/ada4ab}

\bibitem[{{Willott} {et~al.}(2024){Willott}, {Desprez}, {Asada}, {Sarrouh}, {Abraham}, {Brada{\v{c}}}, {Brammer}, {Estrada-Carpenter}, {Iyer}, {Martis}, {Matharu}, {Mowla}, {Muzzin}, {Noirot}, {Sawicki}, {Strait}, {Rihtar{\v{s}}i{\v{c}}}, \& {Withers}}]{Willott24}
{Willott}, C.~J., {Desprez}, G., {Asada}, Y., {et~al.} 2024, \apj, 966, 74, \dodoi{10.3847/1538-4357/ad35bc}

\bibitem[{{Witstok} {et~al.}(2023){Witstok}, {Jones}, {Maiolino}, {Smit}, \& {Schneider}}]{Witstok23}
{Witstok}, J., {Jones}, G.~C., {Maiolino}, R., {Smit}, R., \& {Schneider}, R. 2023, \mnras, 523, 3119, \dodoi{10.1093/mnras/stad1470}

\bibitem[{{Yanagisawa} {et~al.}(2024){Yanagisawa}, {Ouchi}, {Nakajima}, {Harikane}, {Fujimoto}, {Ono}, {Umeda}, {Nakane}, {Yajima}, {Fukushima}, \& {Xu}}]{Yanagisawa24}
{Yanagisawa}, H., {Ouchi}, M., {Nakajima}, K., {et~al.} 2024, arXiv e-prints, arXiv:2411.19893, \dodoi{10.48550/arXiv.2411.19893}

\bibitem[{{Ziparo} {et~al.}(2023){Ziparo}, {Ferrara}, {Sommovigo}, \& {Kohandel}}]{Ziparo23}
{Ziparo}, F., {Ferrara}, A., {Sommovigo}, L., \& {Kohandel}, M. 2023, \mnras, 520, 2445, \dodoi{10.1093/mnras/stad125}

\end{thebibliography}
\end{document}